\documentclass{aa}
 \usepackage{natbib}
\bibpunct{(}{)}{;}{a}{}{,} 
\usepackage[utf8]{inputenc}

\usepackage[T1]{fontenc}
\usepackage{textcomp}
\usepackage{lmodern}
\DeclareUnicodeCharacter{00A0}{ }
\DeclareUnicodeCharacter{202F}{ }
\DeclareUnicodeCharacter{00C1}{\'A}

\usepackage{tabularx}
\usepackage[dvipsnames]{xcolor}
\usepackage{siunitx}
\usepackage{lscape}
\usepackage{amsmath}
\usepackage[autostyle=true]{csquotes}
\usepackage{caption}
\usepackage{graphicx}
\usepackage[colorinlistoftodos]{todonotes}
\usepackage[colorlinks=true, allcolors=black]{hyperref}
\usepackage{caption}
\usepackage[parfill]{parskip}
\usepackage{setspace}
\usepackage{enumerate}
\usepackage{booktabs}
\usepackage{fancyhdr}
\usepackage[english]{babel}
\usepackage{float}
\usepackage{pgfgantt}
\usepackage{footnote}
\usepackage{tablefootnote}
\usepackage{subcaption}
\usepackage{pdflscape}
\usepackage{amssymb}
\usepackage[varg]{txfonts}
\usepackage{orcidlink}

\begin{document} 

   \title{High CO/H$_2$ ratio supports an exocometary origin for a CO-rich debris disk}

\author{
K.D. Smith\inst{1}\,\orcidlink{0000-0002-9274-8715}              \thanks{Email: ksmith9@tcd.ie}
\and
L. Matrà\inst{1}\,\orcidlink{0000-0003-4705-3188}
\and
K. Zhang\inst{2}\,\orcidlink{0000-0002-0661-7517}
\and
A. Brennan\inst{1}\,\orcidlink{0000-0002-7050-0161}
\and
A. M. Hughes\inst{3}\,\orcidlink{0000-0002-4803-6200}
\and
C. Chen\inst{4}$^,$\inst{5}\,\orcidlink{0000-0002-8382-0447}
\and
I. Rebollido\inst{6}\,\orcidlink{0000-0002-4388-6417}
\and
D. Wilner\inst{7}\,\orcidlink{0000-0003-1526-7587}
\and
A. Roberge\inst{8}\,\orcidlink{0000-0002-2989-3725}
\and
S. Redfield\inst{9}\,\orcidlink{0000-0003-3786-3486}
\and
A. Hales\inst{10}\,\orcidlink{0000-0001-5073-2849}
\and
K. \"Oberg\inst{7}\,\orcidlink{0000-0001-8798-1347}
}

   \institute{School of Physics,
              Trinity College Dublin, College Green, Dublin 2, Ireland\\
         \and
         Department of Astronomy, University of Wisconsin-Madison, Madison, WI 53706, USA
         \and 
         Department of Astronomy, Van Vleck Observatory, Wesleyan University, Middletown, CT 06459, USA
         \and
        William H. Miller III Dept. of Physics and Astronomy, John’s Hopkins University, 3400 N. Charles Street, Baltimore, MD 21218, USA
        \and 
        Space Telescope Science Institute, 3700 San Martin Drive, Baltimore, MD 21218, USA
        \and 
        European Space Agency (ESA), European Space Astronomy Centre (ESAC), Camino Bajo del Castillo s/n, 28692 Villanueva de la Cañada, Madrid, Spain
        \and
        Center for Astrophysics, Harvard and Smithsonian, 60 Garden Street, Cambridge, MA 02138-1516, USA
        \and
        NASA Goddard Space Flight Center, Greenbelt, United States
        \and
        Wesleyan University Astronomy Department, Van Vleck Observatory, 96 Foss Hill Drive, Middletown, CT 06459, USA;
        \and
        National Radio Astronomy Observatory, 520 Edgemont Road, Charlottesville, VA 22903-2475, USA
         }

   \date{Received ...; accepted ...}
\abstract
{Over 20 exocometary belts host detectable circumstellar gas, mostly in the form of CO. Two competing theories for its origin have emerged, positing the gas to be primordial or secondary. Primordial gas survives from the belt's parent protoplanetary disk and is therefore H$_2$-rich. Secondary gas is outgassed in-situ by exocomets and is relatively H$_2$-poor. Discriminating between these scenarios has not been possible for belts hosting unexpectedly large quantities of CO.}
{We aim to break this gas origin dichotomy via direct measurement of H$_2$ column densities in two edge-on CO-rich exocometary belts around $\sim$15 Myr-old A-type stars, constraining the $\frac{\text{CO}}{\text{H}_2}$ ratio and CO gas lifetimes. Observing edge-on belts enables rovibrational absorption spectroscopy against the stellar background.}
{We present near-IR CRIRES+ spectra of HD 110058 and HD 131488 which provide the first direct probe of H$_2$ gas in CO-rich exocometary belts. We target the H$_2$ (v=1-0 S(0)) line at 2223.3 nm and the $^{12}$CO $v=2\rightarrow0$ rovibrational lines in the range 2333.8-2335.5 nm and derive constraints on column densities along the line-of-sight to the stars. }
{We strongly detect $^{12}$CO but not H$_2$ in the CRIRES+ spectra. This allows us to place $3\sigma$ lower limits on the $\frac{\text{CO}}{\text{H}_2}$ ratios of $> 1.35 \times 10^{-3}$ and $> 3.09 \times 10^{-5}$ for HD 110058 and HD 131488 respectively. These constraints demonstrate that at least for HD 110058, the exocometary gas is compositionally distinct and significantly H$_2$-poor, compared to the $<10^{-4}$ $\frac{\text{CO}}{\text{H}_2}$ ratios typical of protoplanetary disks. For HD 131488, we further compare the CO photodissociation timescale to the age of the system through simple shielding arguments, finding that we cannot formally rule out a primordial origin; however, we suggest that a more realistic model of CO survival is likely to support a secondary origin for this system as well. Overall, a high $\frac{\text{CO}}{\text{H}_2}$ ratio for HD 110058 indicates that the gas in this CO-rich belt is most likely not primordial in composition, supporting the presence of exocometary gas.}
{}


\titlerunning{High CO/H$_2$ ratios suggest exocometary origin for the gas in a CO-rich host}
\authorrunning{Smith et al.}

   \maketitle

\section{Introduction}
\label{section:1introduction}

Exocometary belts, also known as debris disks, are circumstellar rings composed mainly of dust and rocky/icy planetesimals which have optically thin continuum emission, in contrast to protoplanetary disks \citep[][provide reviews of the field]{pearce_debris_2024, wyatt_extrasolar_2020, hughes_debris_2018}. Exocometary belts are typically detected through the thermal infrared excess they produce over the stellar photosphere. They can be imaged at shorter wavelengths through scattered light from small dust grains with instruments such as the Gemini Planet Imager/Sphere \citep{crotts_uniform_2024, desgrange_dust_2025} and in mm emission with interferometers such as The Atacama Large Millimeter/submillimeter Array (ALMA) \citep{matra_resolved_2025}. Dust in exocometary belts is continually removed by radiation pressure, stellar winds, and/or Poynting-Robinson drag \citep[e.g.][]{strubbe_dust_2006, backman_main-sequence_1993} and replenished by destructive collisional processes between larger grains and planetesimals \citep{dohnanyi_collisional_1969}. Thus, exocometary dust is thought to be secondary in origin, rather than a remnant of the belt's progenitor protoplanetary disk. 

Exocometary belts were initially assumed to be gas-free; however a variety of atomic species \citep[e.g.][]{hobbs_gaseous_1985} as well as CO have been detected in over 20 exocometary systems \citep[e.g.][]{brennan_low_2024, cataldi_primordial_2023, marino_population_2020, kral_predictions_2017, moor_molecular_2017, matra_detection_2017, lieman-sifry_debris_2016, kospal_alma_2013, troutman_ro-vibrational_2011, dent_co_2005, zuckerman_inhibition_1995}. Thus far CO is the only molecule detected in exocometary gas due to its resistance to photodissociation, chemical stability, and easy detectability \citep{visser_photodissociation_2009}. CO has strong rovibronic transitions, and the typically cold gas present in exocometary belts is detectable in UV rovibronic absorption spectra by the \textit{Hubble Space Telescope} (\textit{HST}) \citep[e.g.][]{roberge_high-resolution_2000}, rotational transitions in emission detectable in mm with ALMA \citep[e.g.][]{matra_detection_2017, cataldi_primordial_2023, rebollido_search_2022} and IR absorption from rovibrational transitions as detected using the Gemini South Telescope and the NASA Infrared Telescope Facility around $\beta$ Pictoris \citep{troutman_ro-vibrational_2011}. 

CO emission from rotational transitions observed with ALMA has shown that exocometary belts can be separated into 2 categories of: CO-rich ($M_{\mathrm{CO}} \gtrsim 10^{-3} M_{\bigoplus}$) and CO-poor ($M_{\mathrm{CO}}\lesssim 10^{-3} M_{\bigoplus}$) \citep{marino_population_2020}. This bimodal distribution of exocometary CO masses can be seen in \citet{cataldi_primordial_2023}, with CO masses measured directly through $^{12}$C$^{16}$O lines (if optically thin) or through $^{13}$C$^{16}$O or $^{12}$C$^{18}$O measurements (if optically thick and assuming interstellar isotopologue ratios). This gas could be primordial if persisting from the parent protoplanetary disk \citep{nakatani_primordial_2023, kospal_alma_2013} or secondary if produced in-situ by outgassing from exocomets \citep{zuckerman_40_2012}.

When determining the origin of the CO observed in exocometary belts, it is important to consider the role of UV shielding of CO by self-shielding, CI, and H$_{2}$, which attenuates UV photons and lengthens the time taken for CO to photodissociate \citep{heays_photodissociation_2017, visser_photodissociation_2009}.  Unlike in protoplanetary disks, the dust in exocometary belts is optically thin at all wavelengths \citep[e.g][]{matra_empirical_2018}. Thus, it cannot provide circumstellar gas with substantial shielding from UV radiation due to the host star and interstellar radiation field (ISRF) with the expected lifetime of an unshielded CO molecule in the ISRF being $\sim$ 130 years \citep{heays_photodissociation_2017}. If the gas in an exocometary belt has a photodissociation time shorter than the age of the system, this implies the gas is secondary, as it likely is being replenished by comets in the belt. Furthermore, CO-rich belts require shielding to increase the longevity of outgassed CO as the rate of outgassing would be unreasonably high to explain the masses observed if the CO were destroyed on short timescales \citep{kospal_alma_2013, kral_imaging_2019}.

If the gas is primordial, the abundant H$_2$ inherited from the protoplanetary disk could provide shielding \citep{kospal_alma_2013}. If the gas is secondary, CO can be replenished by outgassing from exocomets in the belt \citep{zuckerman_40_2012}, and shielding could be provided by CO itself (self-shielding) and/or atomic carbon \citep{matra_exocometary_2017, kral_imaging_2019}, which could be produced by photodissociation of carbon-bearing parent molecules such as CO but also yet unseen species such as CO$_2$ and CH$_4$. Comparing the CO photodissociation timescale to the system age can be used to determine the gas origin scenario in CO bearing debris disks. Timescales longer than the system age allow for a primordial origin and shorter implies secondary gas.

In CO-rich debris disks, assuming a $\frac{\text{CO}}{\text{H}_2}$ abundance ratio typical of protoplanetary disks leads to sufficient H$_2$ to shield CO for several Myr, close to the system age \citep[e.g.][]{kospal_alma_2013}. However, the same assumption for CO-poor belts leads to insufficient H2 to shield CO over the system age \citep[e.g][]{marino_exocometary_2016, matra_detection_2017, marino_scattering_2018, matra_ubiquity_2019, kral_imaging_2019}. This implies that gas in CO-poor systems, from younger systems like $\beta$ Pictoris \citep{matra_exocometary_2017} to older systems like Fomalhaut \citep{matra_detection_2017}  and $\eta$ Corvi \citep{marino_exocometary_2016}, is most likely secondary. This was corroborated by upper limits on the H$_2$ density in the $\beta$ Pictoris disk obtained from non-local thermodynamic equilibrium (non-LTE) modelling of the CO excitation \citep{matra_exocometary_2017}.

In CO-rich belts, on the other hand, the origin of the gas remains an open question. These disks have CO masses comparable with the lower end observed in Herbig-Ae protoplanetary disks but around stars that are several Myr to a few 10s of Myr old \citep{moor_molecular_2017}.  If the gas in these CO-rich disks is primordial, we would expect a large amount of H$_2$, the most abundant molecular species in primordial gas. In protoplanetary disks, the gas phase CO-to-H$_2$ ratio appears to decrease over time, producing $\frac{\text{CO}}{\text{H}_2}$ ratios between $10^{-4}$ and $10^{-6}$ (see Figure \ref{fig:col_den_ratios}, Table \ref{tab:co_h2_ratios} and \citet{bergin_determination_2018, zhang_rapid_2020}). This is because CO abundances start close to ISM-like, but over time CO can be processed into other  carbon-carriers (such as CO$_2$ and CH$_3$OH) and at the same time CO can be sequestered onto icy grains as they grow and settle to the mid-plane \citep{schwarz_unlocking_2018, bosman_co_2018, krijt_transport_2018, krijt_co_2020}.

On the other hand, for second generation gas, we may look to Solar System comets for comparison. Here, $\frac{\text{CO}}{\text{H}_2}$ ratios as high as 0.73 have been observed as H$_2$ is mainly produced as a product of H$_2$O photodissociation \citep{bockelee-morvan_composition_2017, feldman_far_2002}. No molecules other than CO have thus far been observed in debris disks, thus it is unconfirmed whether exocometary ices should have compositions comparable to solar system cometary ice, but it is reasonable to expect a $\frac{\text{CO}}{\text{H}_2}$ abundance ratio higher than 10$^{-4}$ in secondary gas.

Some constraints on the presence of H$_2$ in exocometary belts are present in the literature. For example, using \textit{HST} \citet{lecavelier_des_etangs_deficiency_2001} constrained upper limits for the column density of H$_2$ along the line of sight to the edge-on, CO-poor $\beta$ Pictoris disk to be < 0.1 M$_\oplus$ and combined this with a CO column density \citep{roberge_high-resolution_2000} to find a $\frac{\text{CO}}{\text{H}_2}$ ratio lower limit of $6\times 10 ^{-4}$, higher than expected of primordial gas. Their conclusion is in agreement with the conclusions of \citet{matra_exocometary_2017} that a primordial origin scenario for the gas in $\beta$ Pictoris is unlikely as there is insufficient H$_2$ present to provide shielding for the lifetime of the system. H$_2$ can also be indirectly constrained by gas kinematics/distribution. \citet{fernandez_braking_2006} use the breaking experienced by ionised metallic gas due to unseen neutral H$_2$ to conclude that the gas observed in $\beta$ Pictoris is likely secondary and produced in-situ due to evaporation of dust grains in the belt. HI Lyman $\alpha$ emission was detected by \citet{wilson_first_2017}, but given that the hydrogen content of the disk is less than solar abundances, the authors conclude the HI likely originates from the dissociation of H$_2$O released by cometary ices rather than H$_2$ originating from the protoplanetary disk. 
\citet{hughes_radial_2017} use the scale height of the gas in 49 Ceti to infer an upper limit for H$_2$, via the mean molecular weight of the gas required by models to reproduce the observed scale height for CO. However, as pointed out by \citet{marino_vertical_2022} this is only valid if the CO scale height traces the scale height of the bulk of the gas, which may not be the case.

Hot H$_2$ emission has also been detected in systems bearing debris disks such as TWA 7 \citep{flagg_detection_2021} and AU Mic \citep{flagg_mysterious_2022}, where the latter has no evidence for CO \citep{cronin-coltsmann_alma_2023}.  
It is unclear whether this hot H$_2$ could originate from star spots or an inner disk of gas accreting onto the star, unrelated to the outer disks. Other works have attempted to detect warm $\text{H}_2$ with Spitzer, but these observations lacked the SNR/sensitivity to confirm or refute a primordial origin scenario \citep[]{chen_dust_2007, kospal_alma_2013}.

In this work, we aim to directly probe the column density of H$_2$ within the CO-rich exocometary belts HD 110058 and HD 131488 for the first time and in doing so establish the CO origin. We expect any exocometary $\text{H}_2$ to be co-located with and similar temperature as the CO detected through UV absorption spectroscopy in both disks by \citet{brennan_low_2024} and in IR in this work. However it is worth noting that H$_2$ should have a larger vertical extent than CO due to CO being more readily photodissociated at the disk's surface and it's lower mean molecular weight \citep{marino_population_2020, hughes_radial_2017}. Both H$_2$ and CO should be cold and produce narrow absorption lines from their ground vibrational energy levels.  We focus on absorption spectroscopy in edge-on belts to use the star as a bright background continuum, and maximise the column density of gas along the line of sight to it.  
We identified the H$_2$ v=1-0 S(0) line at 2223 nm as the most promising transition probing rovibrational absorption from the ground level of the H$_2$ molecule. This is because, assuming the H$_2$ gas in these systems has a similar temperature to the CO and is in LTE (predominantly collisionally excited), this line would be the strongest IR transition. 

We used the newly upgraded, high-resolution CRIRES+ instrument on the VLT to observe the edge-on, CO-rich disks around HD 110058 and HD 131488.  With a spectral resolution $R \sim 100000$ \citep{leibundgut_science_2022}, CRIRES+ is ideal for detecting line of sight absorption due to CO and H$_2$. Our observations cover multiple v=2-0 rovibrational $^{12}$CO lines. Absorption from CO in edge-on debris disks has been successfully observed by \citet{troutman_ro-vibrational_2011}, who reported low-$J$ CO absorption from the $\beta$~Pictoris disk using IRTF/CSHELL. \citet{worthen_vertical_2024} reported a non-detection of CO in absorption toward the edge-on HD 32297 disk using IRTF/iSHELL which was used to estimate the scale height of the disk and place an upper limit on the line-of-sight CO column density. The two disks in this study were measured in absorption in the UV using \textit{Hubble} by \citet{brennan_low_2024}. Our CRIRES+ observations allow for CO column density and temperature estimates for both HD 110058 and HD 131488.

Stellar and disk properties for HD 110058 and HD 131488 are detailed in Table \ref{table:stellar_params}. Both are A-type stars \citep[the most common spectral type for CO-rich belts,][]{moor_molecular_2017} and thus provide bright stellar continua against which we searched for absorption. The disks in these systems are also very close to edge-on with inclinations of $85.5^{+2.5}_{-7.2}$$^{\circ}$ and $82\pm3$$^{\circ}$ respectively \citep{moor_molecular_2017, hales_alma_2022}. We chose these two young  ($\sim15$ Myr), CO-rich, edge-on belts as they have the highest CO column densities detected in absorption to date \citep{brennan_low_2024}.

\begin{table*}
    \caption{Stellar and Disk Parameters for HD 131488 and HD 110058.}
    \label{table:stellar_params}      
    \centering          
    \begin{tabular}{l c c c c}    
        \hline\hline       
        Parameter & HD 131488 & References & HD 110058 & References \\ 
        \hline                    
        Spectral Type & A1 V & [1] & A0 V & [2] \\  
        R.A. (J2000) & 14 55 08.03 & [1] & 12 39 46.14 & [2] \\  
        Decl. (J2000) & -41 07 13.3 & [1] & -49 11 55.84 & [2] \\  
        K-band Magnitude & 7.803 & [3] & 7.583 & [3] \\  
        Radial Velocity (km s$^{-1}$) & 5.97$^{+0.05}_{-0.05}$ & [8] & 12.53$^{+0.01}_{-0.01}$ & [8] \\  
        Distance (pc) & 152$^{+3.2}_{-0.2}$ & [4] & 130$^{+2}_{-0.2}$ & [4] \\  
        Association & Upper Centaurus Lupus & [1] & Lower Centaurus Crux & [2] \\  
        Age & $\sim$16 Myr & [1] & $\sim$17 Myr & [2] \\  
        Log(g) & 3.80$^{+0.38}_{-0.16}$ & [4] & 3.54$^{+0.41}_{-0.01}$ & [4] \\  
        Stellar Radius (R$_\odot$) & 1.60$^{+0.03}_{-0.02}$ & [4] & 1.55$^{+0.02}_{-0.04}$ & [4] \\  
        Stellar Mass (M$_\odot$) & 1.8 & [5] & 1.84 & [2] \\  
        Disk Inclination ($^\circ$) & $82^{+3}_{-3}$ & [6] & 85.5$^{+2.5}_{-7.2}$ & [2] \\  
        $R_{\rm in}$ (au) & $35^{+11}_{-11}$ & [7] & $7.4^{+2.2}_{-7.3}$ & [2] \\  
        $R_{\rm out}$ (au) & $140^{+11}_{-11}$ & [7] & $80 \pm 12$ & [2] \\  
        log($N_{\text{CO}}$ (cm$^{-2}$)) & 18.1$^{+0.2}_{-0.1}$ & [4] & 19.7$^{+0.1}_{-0.1}$ & [4] \\  
        T$_{\text{gas}}$ (K) & 141$^{+103}_{-55}$ & [8] & 133.0$^{+19}_{-18}$ & [8] \\  
        \hline                  
    \end{tabular}
    \tablebib{
        [1]~\citet{melis_copious_2013}; [2] \citet{hales_alma_2022}; [3] \citet{cutri_vizier_2003}; 
        [4] \citet{brennan_low_2024}; [5] \citet{matra_empirical_2018}; [6] \citet{moor_molecular_2017}; 
        [7] \citet{smirnov-pinchukov_lack_2022}; [8] This work. Note: T$_{\text{gas}}$ is the kinetic temperature from the $^{12}$CO observations summarised in Appendix \ref{appendix:MCMC}. It is only the same as the kinetic temperature if the system is in LTE and the H$_2$ is colocated with the CO. $R_{\rm in}$ and $R_{\rm out}$ are the radii for the CO disk while the disk inclination is derived from the dust disk. The radial velocity is in the barycentric frame.
    }
\end{table*}

In Section \ref{section:2observations} we describe the CRIRES+ observations and data reduction. In Section \ref{section:model} we describe the modelling of H$_2$ and $^{12}$CO for the two disks and the resulting constraints on the $\frac{\text{CO}}{\text{H}_2}$ ratio. In Section \ref{section:4Discussion} we compare this ratio with observations of protoplanetary disks and describe how this impacts our understanding of gas detections in CO-rich exocometary belts. In Section \ref{section:5conclusions} we conclude with a summary of our findings.

\section{Observations and data reduction \label{section:2observations}}
\subsection{CRIRES+ observations}

Using CRIRES+, the high-resolution near infrared spectrograph on the VLT \citep{dorn_crires_2023}, we observed HD 110058 and HD 131488 over two nights each, as detailed in Table \ref{tab:observations}. We observed in the K-band with the K2148 filter, which has non-contiguous wavelength orders from 1904 nm to 2452 nm. Our observation dates were chosen such that the H$_2$ v=1-0 S(0) rovibrational absorption line at 2223 nm was shifted away from nearby weak tellurics, mainly arising from atmospheric water as can be seen in Figure \ref{fig:molecfot}. Our observation windows also covered the weaker H$_2$ v=1-0 Q(1) transition at 2406.6 nm. If detected, this transition could allow us to derive the excitation temperature of the H$_2$ gas. Also within the range of the data are the v=2-0 bands from $^{12}$CO, $^{13}$CO, and C$^{18}$O. The latter 2 isotopologues will be examined in an upcoming work.

\begin{table*}
    \caption{Overview of VLT-CRIRES+ observations of HD 110058 and HD 131488.}
    \label{tab:observations}
    \centering
    \begin{tabular}{c c c c c c c c c}
        \hline\hline
        Target & Programme ID & Night & Exp. time & Obs. mode & Slit & AO loop & Wavelength setting & Airmass \\ 
        \hline                    
        HD 110058 & 110.248L.001 & 25/03/2023 & $12\times240$s & Nodding & 0.2" & Closed & K2148 & 1.134 \\  
        HD 110058 & 110.248L.001 & 08/03/2024 & $12\times240$s & Nodding & 0.2" & Closed & K2148 & 1.106 \\  
        HD 131488 & 111.255A.001 & 08/05/2023 & $24\times120$s & Nodding & 0.2" & Closed & K2148 & 1.168 \\  
        HD 131488 & 111.255A.001 & 11/05/2023 & $24\times120$s & Nodding & 0.2" & Closed & K2148 & 1.122 \\  
        \hline                  
    \end{tabular}
    \tablefoot{
        Exposure time is expressed as number of detector integrations (NDIT) $\times$ detector integration time (DIT).
    }
\end{table*}

\subsection{CRIRES+ data reduction and calibration} \label{subsection:reduction}
\begin{figure}
    \centering
    \includegraphics[width=\linewidth]{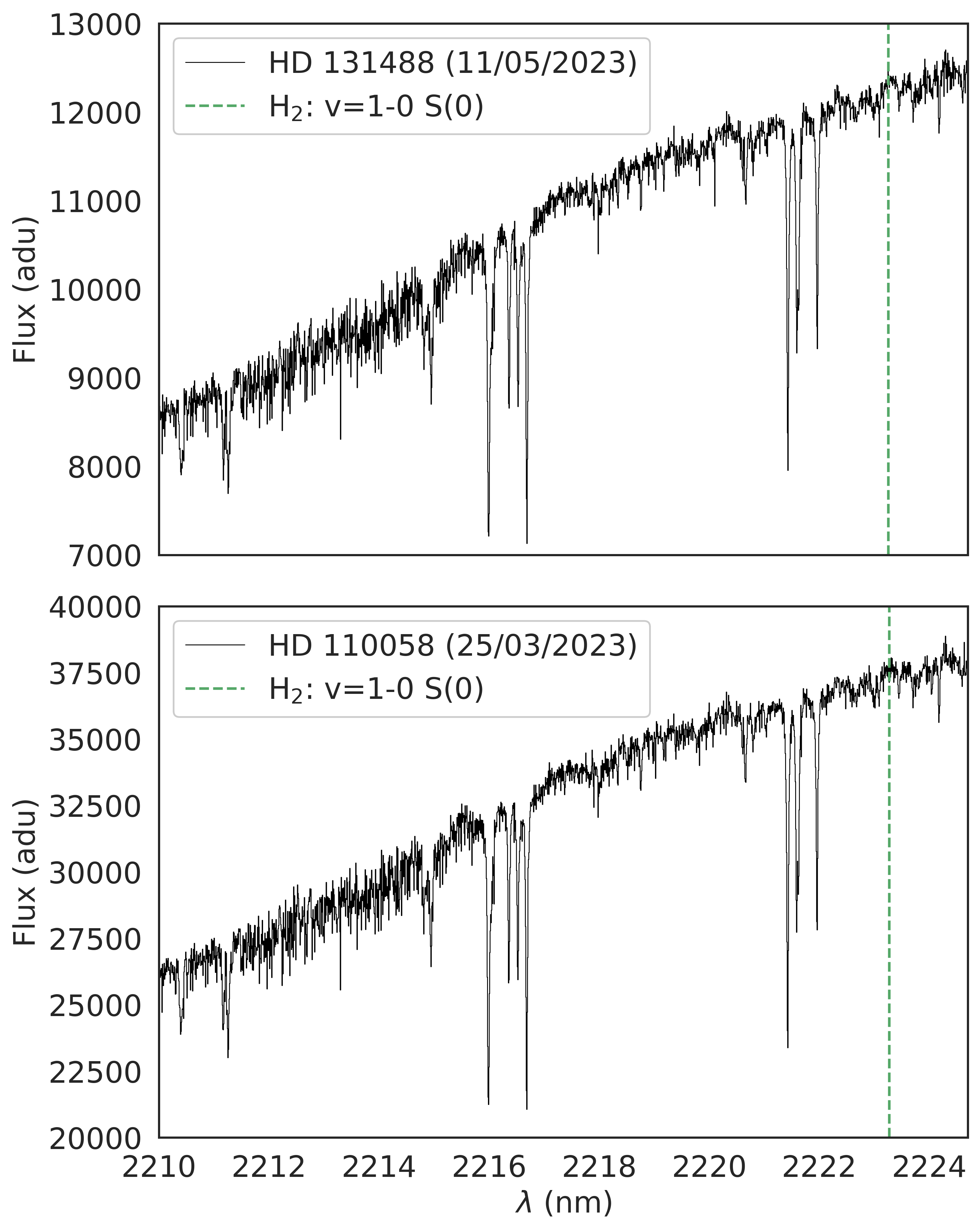}
    \caption{Reduced and extracted 1D spectra (black lines) for a portion of the detector-order containing our H$_2$ line of interest for HD 131488 on 11/05/23 (top) and HD 110058 on 25/03/23 (bottom). The spectra are not flux calibrated and as such the absolute values of the y-axis in units of analogue-to-digital units (adu) are not astrophysically meaningful. The spectra are not blaze corrected and thus the flux tends to rise with wavelength. The vertical green line denotes the expected location of the H$_2$ transition of interest. The spectra are in the observatory frame and the H$_2$ location has been shifted to account for the Doppler shift of the star compared to the observatory frame.}
    \label{fig:raw_observations}
\end{figure}

We used ESOREX and the CR2RES pipeline recipes detailed in the ESO CRIRES+ pipeline user manual (version 1.3.0) to reduce the raw detector images. The standard reduction for nodding observations shown in Figure 4.4 of this manual was the basis of our reduction. High-SNR flats for CRIRES+ are taken once a month, and were used instead of the default supplied flats to improve the final SNR. We performed wavelength calibration with the Uranium-Neon lamp and Fabry-Pérot etalon. After correcting for detector non-linearity, dark current, and then carrying out flat fielding and wavelength calibration, we extracted a 1D spectrum for each of the two nodding positions. The nodding positions both contained the star but at different locations in the detector to account for localised detector systematics. A subset of the extracted data (with the two nodding positions combined) covering the wavelength of the H$_2$ line of interest is shown in Figure \ref{fig:raw_observations}. This data is yet to be corrected for tellurics or blaze; as such it has arbitrary flux units and displays an upward trend with wavelength.

Our planned analysis required achieving a high-SNR and modelling the line-to-continuum ratio in the continuum normalised spectrum as detailed in Section \ref{section:model}. Since flux calibration is not necessary for this goal, a standard star observation was not needed. We performed telluric correction using the \texttt{molecfit} software, version 4.3.1 \citep{smette_molecfit_2015} which has been shown to perform as well or better than corrections using standard stars on CRIRES data \citep{ulmer-moll_telluric_2019}. We performed telluric correction and additional telluric-based wavelength calibration on subsets of the data around lines of interest. For more details on the wavelength calibration and why it was useful for our data, see Appendices \ref{appendix:molecfit} and \ref{appendix:superresolution}. Figures \ref{fig:molecfot} and \ref{fig:12CO_telluric} show example spectra with telluric fits with the most prominent, yet relatively weak tellurics arising from atmospheric water, while corrected spectra are shown in Figures \ref{fig:hd131488_radis_wide}, \ref{fig:hd110058_CO_model}, and \ref{fig:hd131488_CO_model}. We note that by design, the H$_2$ line under investigation does not overlap with any telluric lines, thus removing tellurics does not significantly affect the spectrum at or near the line. Telluric removal was also carried out for each nodding position/observing day combination for the $^{12}$CO lines in spectral regions covering the v=2-0 rovibrational band from 2318.9 to 2333.9 nm and 2335.455 to 2350.0 nm and for the H$_2$ v=1-0 S(0) line from 2220 to 2224.5 nm. The nodding positions were modelled separately to allow for improved wavelength calibration described in Appendix \ref{appendix:superresolution}. An advantage of using \texttt{molecfit} is that in addition to telluric removal, it also returns a spectral resolution for the observations, specific to each detector-order, nodding position and observing day combination.

The average resolution over each spectrum in the region corrected by \texttt{molecfit} near the H$_2$ line, was measured to be $\text{R} = 126000 \pm 9000$ for HD 131488 and $\text{R} = 130000 \pm 7000$ for HD 110058. For the regions containing the $^{12}$CO, the resolution was measured to be $\text{R} = 115000 \pm 9000$ and $\text{R} = 121000 \pm 9000$ for HD 131488 and HD 110058 respectively. The resolution varies with the spectral region under consideration, and the weather. CRIRES+ typically has a resolution of 100000, but because the stars did not fill the slit width, we are in the superresolution regime described in Appendix \ref{appendix:superresolution}. This leads to offsets in wavelength between nodding positions. Thus, we did not use the final CRIRES nodding corrected spectra. Instead we telluric corrected the spectra for each nodding position separately and used the tellurics to ensure each spectrum was correctly wavelength calibrated before modelling.

To identify the $\text{H}_2$ and CO lines at their rest frequencies, the reduced telluric free spectra for each observing night were Doppler shifted to barycentric frame and then to the rest frame of the star using the system radial velocities. The barycentric velocity correction is calculated using Astropy's SkyCoord class \citep{the_astropy_collaboration_astropy_2022}, whereas radial (systemic) velocities for each system were determined through modelling $^{12}$CO in the CRIRES+ data (see Section \ref{section:model} and Table \ref{table:stellar_params}). The spectra \texttt{molecfit} returns are telluric-corrected but not continuum-normalised. We divide the telluric-corrected spectrum with a running median with a large window size to produce continuum normalised spectra. 

Our final normalised spectra are heteroscedastic, as across each order the flux of the star changes, and telluric correction produces regions of higher/lower variance. To calculate the variance and therefore an uncertainty per pixel empirically, we therefore masked regions around the CO and H$_2$ lines of interest and calculated rolling variances with a 20-pixel-wide window, chosen to best capture local variations in variance around the target lines. For the masked pixel locations around these lines, we then linearly interpolated the variance from neighbouring pixel regions. The final products used for modelling are 4 spectra of normalised intensity and their empirically-determined uncertainty per target, one for each of two nodding positions and two observing dates. We then median averaged the nods and dates to produce the spectra in Figures \ref{fig:hd131488_radis_wide}, \ref{fig:hd110058_CO_model}, and \ref{fig:hd131488_CO_model}.

\section{Results and modelling} 
\label{section:model}

\begin{figure}
    \centering
    \includegraphics[width=\linewidth]{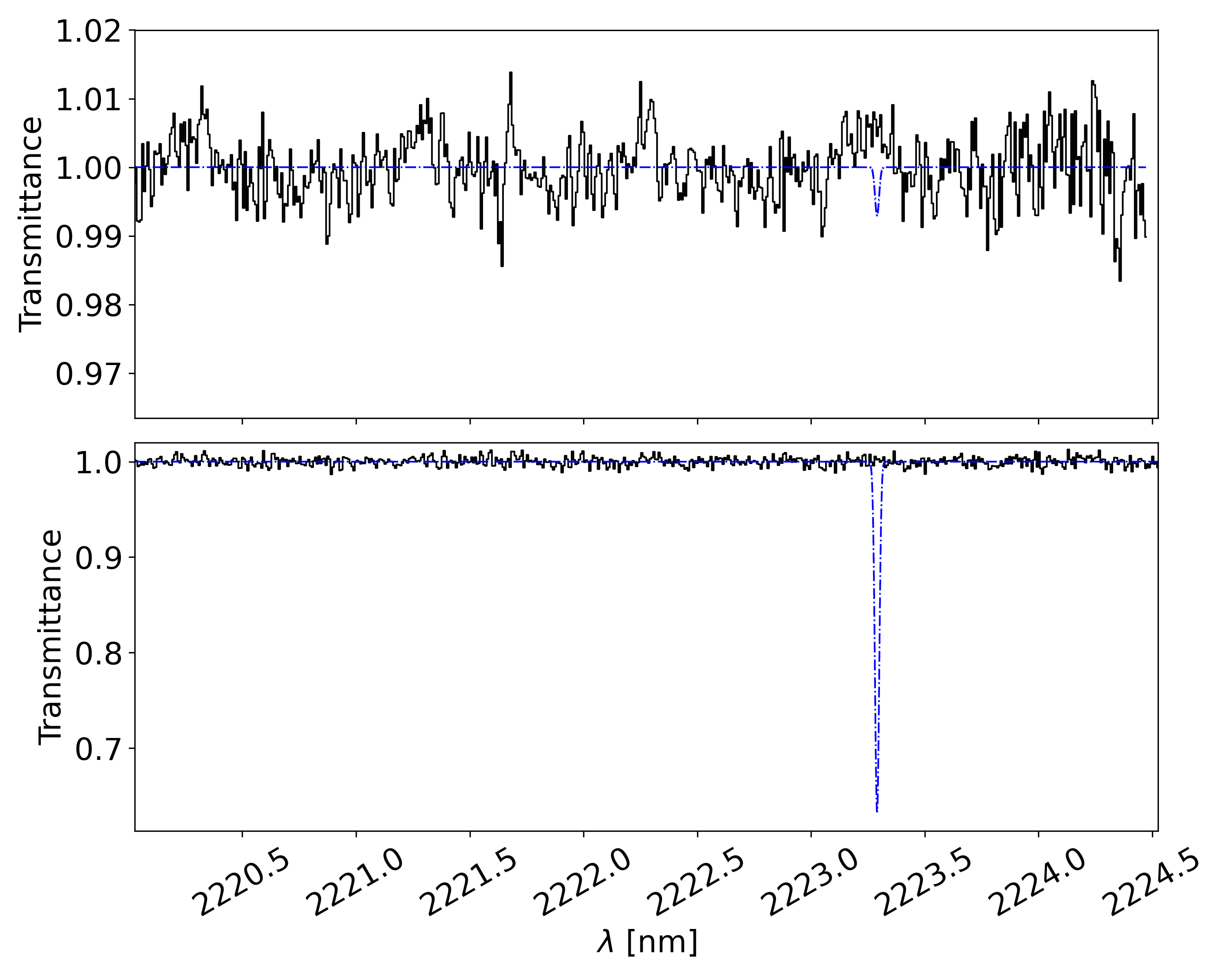}
    \caption{In black are the median, telluric corrected, normalised spectra of HD 131488 (top) and HD 110058 (bottom). In blue is a model of the H$_2$ v=1-0 S(0) line assuming a temperature corresponding to the CO kinetic temperature we derive from our results and an H$_2$ column density corresponding to an ISM-like, primordial $\frac{\text{CO}}{\text{H}_2}$ ratio of $10^{-4}$.}
    \label{fig:hd131488_radis_wide}
\end{figure}

The final normalised spectra for each star are displayed in Figures \ref{fig:hd131488_radis_wide}, \ref{fig:hd110058_CO_model}, and \ref{fig:hd131488_CO_model}, with wavelengths in the rest frame of the star. For $\text{H}_2$, if detectable, we expect circumstellar gas to produce an absorption line at the v=1-0 S(0) transition wavelength of 2223.29 nm \citep[from the HITRAN database,][]{gordon_hitran2020_2022}. No statistically significant detection of H$_2$ was found for either system. Additionally, as expected for cold $\text{H}_2$ where the ground state should be the most populated energy level, the weaker 2406.6 nm absorption line arising from the H$_2$ v=1-0 Q(1) transition was not detected.

We detect $^{12}$CO absorption at a radial velocity consistent with that observed in the UV for HD 110058 by \citet{brennan_low_2024}. The radial velocity for HD 131488 differs from the one reported in \citet{brennan_low_2024} (they report a radial velocity of $4.8^{+0.1}_{-0.1}$ km s$^{-1}$ whereas we find the radial velocity to be $5.97^{+0.05}_{-0.05}$ km s$^{-1}$), but it is consistent with the CO radial velocity observed in emission with ALMA by \citet{macmanamon_alma_2026} as part of the ARKS program. For HD 110058, we detect the v=2-0, J=15-14 to J=1-0 and for HD 131488 we detection the v=2-0, J=8-7 to J=1-0 transitions. More CO is present in HD 110058, and thus the absorption lines are deeper, reflected in the axis scaling of Figures \ref{fig:hd110058_CO_model} and \ref{fig:hd131488_CO_model}. Additionally, the excitation temperature for HD 110058 is higher, allowing more high-J transitions to be populated.

\subsection{Modelling the reduced CRIRES+ spectra}
We modelled H$_2$ absorption in the normalised spectra using RADIS \citep{pannier_radis_2019}. RADIS relies on the HITRAN line list \citep{gordon_hitran2020_2022} obtained using the HITRAN Application Programming Interface \citep{kochanov_hitran_2016}, and partition functions described in \citet{gamache_total_2021}. Given the non-detection, for H$_2$ we assumed the gas is in LTE and that it can be approximated as having a single column density and temperature along the line of sight to the star. On the other hand, for CO we found LTE not to be a sufficiently good description of the data and we therefore adopted a simplified non-LTE with a different kinetic temperature (setting the intrinsic line width through Doppler broadening) from the excitation temperature (setting the ratio of the populations of the various rotational/vibrational levels via Boltzmann distributions, and therefore the relative strength of our detected lines). This is consistent with the findings and the modelling of UV rovibronic lines in the same systems \citep{brennan_low_2024}. Although we cannot determine whether H$_2$ is in LTE without more information about the collider densities, the H$_2$ (v=1-0 S(0)) transition has an Einstein A of $2.524\times 10^{-7} \text{s}^{-1}$ while the CO $^{12}$CO $v=2\rightarrow0$ transitions have Einstein A coefficients on the order of $10^{-1} \text{s}^{-1}$. Meanwhile the collisional rate coefficients found in \citet{thi_radiation_2013, quemener_quantum_2009} are of comparable orders of magnitude. This means the H$_2$ transition has a lower critical density and is therefore more likely to be in LTE than the detected CO. 

For a fixed wavelength range RADIS produces a normalised absorption spectrum dependent on temperature (excitation and/or kinetic) and column density, and accounting for optical depth effects. The model spectra are convolved with a Gaussian with a FWHM equal to the spectral resolution of our observations found using the best-fit models from \texttt{molecfit}. A radial velocity shift was also left as a free parameter. This gives us a final convolved, normalised absorption model for the CO and H$_2$ rovibrational transitions for any combination of kinetic/excitation temperature, column density, and radial velocity which can be compared with our normalised spectra for each star. 

We assessed the goodness of fit for each set of model parameters with a log Gaussian likelihood ($\ln{\mathcal{L}}$) function:
\begin{align}
\ln{\mathcal{L}} = -\frac{1}{2}\sum_{i=0}^{n} \left( \frac{R_i^2}{\sigma_i^2} + \mathrm{ln}(2\pi\sigma_i^2) \right)
\end{align}
where $R_i$ are the residuals for the i$^\text{th}$ spectral pixel in the normalised spectrum.
The uncertainties ($\sigma_i$) for the i$^\text{th}$ pixel  are derived empirically as described in Section \ref{subsection:reduction}.

We use a Markov-Chain Monte Carlo (MCMC) method implemented in python's emcee package \citep{foreman-mackey_emcee_2013} to sample the posterior probability distribution. We used uniform priors given in Table \ref{tab:MCMC_priors} for each parameter.

\subsection{Results \label{section:3results}}
The posterior probability distributions from the MCMC fit are shown in Figures  \ref{fig:hd131488_posterior}, \ref{fig:hd110058_posterior}, \ref{fig:hd131488_CO_posterior}, and \ref{fig:hd110058_CO_posterior}. There is clear degeneracy between the H$_2$ temperature and column density which cannot be broken with a non-detection. For larger temperatures, larger column densities are allowed as Doppler broadening causes broader, shallower lines which are harder to detect. When marginalised over temperature, we obtain column density upper limits. The column density posterior is close to a uniform distribution for low column densities as small amounts of H$_2$ would all equally result in signals below the noise level of the data. Larger column densities would produce larger signals reaching detectability and thus these models are less likely to describe our observations. 

\begin{figure}
    \centering
    \includegraphics[width=\linewidth]{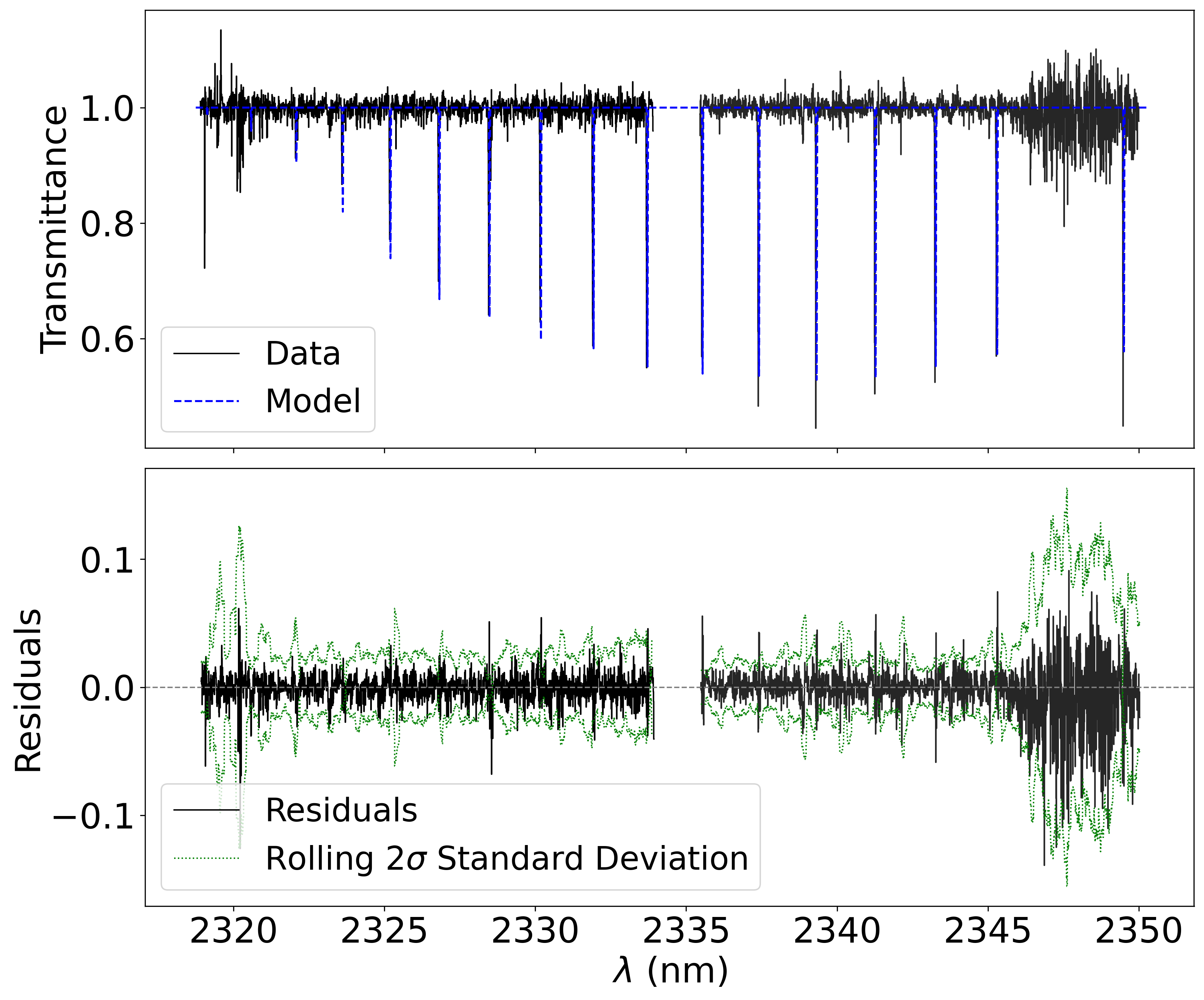}
    \caption{The best fit (highest log posterior probability) $^{12}$CO models, normalised HD 110058 data, and residuals. The top figure shows the HD 110058 $^{12}$CO data (black solid line) overlaid with the best-fit model (blue dotted line). The bottom figure shows the $^{12}$CO residuals (black solid line) and  $2\sigma$ calculated to be twice the rolling standard deviation (green dashed line).}
    \label{fig:hd110058_CO_model}
\end{figure}

\begin{figure}
    \centering
    \includegraphics[width=\linewidth]{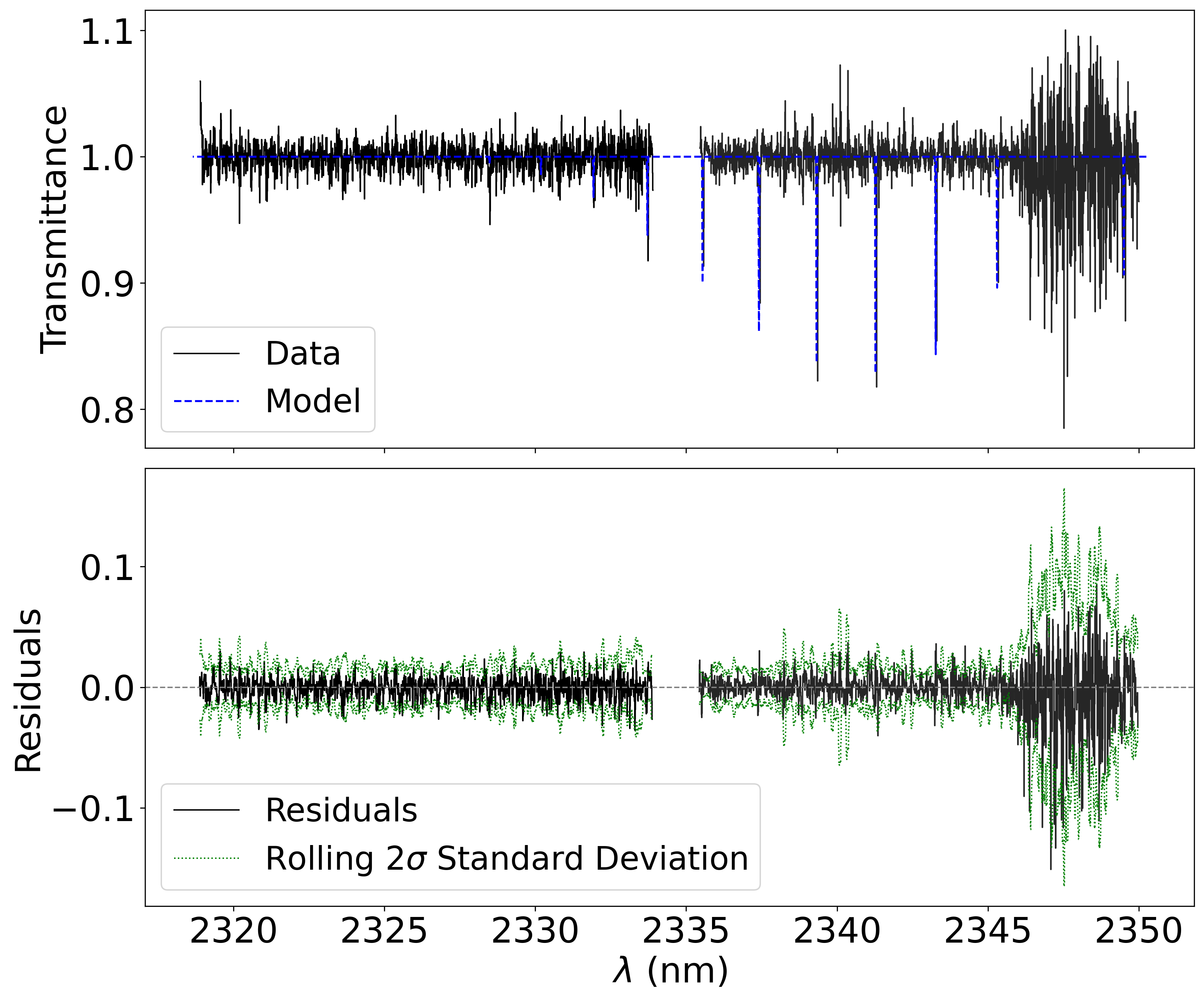}
    \caption{The best fit (highest log posterior probability) $^{12}$CO models, normalised HD 131488 data, and residuals. The top figure shows the HD 131488 $^{12}$CO data (black solid line) overlaid with the best-fit model (blue dotted line). The bottom figure shows the $^{12}$CO residuals (black solid line) and $2\sigma$ calculated to be twice the rolling standard deviation (green dashed line).}
    \label{fig:hd131488_CO_model}
\end{figure}

$^{12}$CO was detected in both systems as shown in Figure \ref{fig:hd110058_CO_model} and \ref{fig:hd131488_CO_model}. We used the same modelling approach as H$_2$, but the detection of the CO band allows us to constrain more parameters, including the radial velocity which is not constrained for H$_2$ but is used to create Figure \ref{fig:hd131488_radis_wide} with the expected position of the H$_2$ line.

The best-fit $^{12}$CO models for each star are shown in Figure \ref{fig:hd110058_CO_model} and \ref{fig:hd131488_CO_model}. Both systems have column densities and excitation temperatures that are well constrained and consistent with those found by \citet{brennan_low_2024}. Our kinetic temperatures are less well constrained given the lines are individually unresolved or at best marginally resolved. The best-fit values are consistent with the more constraining UV results at the 3$\sigma$ level.  The excitation temperatures we derive through MCMC fits (see Figures \ref{fig:hd131488_CO_posterior} and \ref{fig:hd110058_CO_posterior}) are consistent with those derived in the UV by \citet{brennan_low_2024}.

The best-fit parameters (highest posterior probability) for CO are given in Table \ref{tab:MCMC_results} while the posterior distributions are shown in Figures \ref{fig:hd131488_CO_posterior} and \ref{fig:hd110058_CO_posterior} and the best-fit model/residuals are shown in Figures \ref{fig:hd110058_CO_model} and \ref{fig:hd131488_CO_model}. The posterior distributions in Appendix \ref{appendix:MCMC} indicate that the column densities are constrained by the model. We find that our best-fit models are very good at representing the data, with some 3$\sigma$ residuals remaining around some optically thick CO lines in HD 110058 as shown in Figures \ref{fig:hd110058_CO_model} and \ref{fig:hd131488_CO_model} . These residuals could be due to imperfect telluric correction, and/or the assumed perfectly Gaussian shape of the instrumental spectral response, and/or model assumptions such as the assumption that the line of sight CO column can be represented by one temperature and likely don't significantly impact our conclusions.

To calculate the $\frac{\text{CO}}{\text{H}_2}$ ratio lower limits, we assume that the H$_2$ shares the same temperature as the CO kinetic temperature derived from our observations (as described below), which is reasonable if CO and any hypothetical H$_2$ are co-located. The choice of kinetic rather than excitation temperatures is conservative with regards to how much unseen H$_2$ could be present as the colder excitation temperatures derived for CO would lead to lower column densities.

Given the H$_2$ column density upper limit, and the CO detection, we can combine their column densities to derive a $\frac{\text{CO}}{\text{H}_2}$ ratio lower limit for the gas in both systems. As the column densities for both H$_2$ and CO are correlated with the gas temperatures, we paired each CO column density sample starting from the highest log posterior probability sample with a H$_2$ column density at the closest kinetic temperature. This gave us a representative $\frac{\text{CO}}{\text{H}_2}$ probability distribution accounting for the temperature correlations. This is comparable to if we had fit CO and H$_2$ simultaneously and enforced both having the same temperature. We then took the $99.7^{\text{th}}$ percentile of this distribution as the $3\sigma$ lower limit for each system, reported in Table \ref{tab:MCMC_results}.

\begin{table}
\caption{CRIRES+ CO/H$_2$ modelling results.}
\label{tab:MCMC_results}
\centering
\begin{tabular}{c c c}
\hline\hline
Parameter & HD 110058 & HD 131488 \\
\hline
log(N$_{\rm CO}$ (cm$^{-2}$)) 
& $20.0^{+0.3}_{-0.2}$ 
& $18.05^{+0.06}_{-0.04}$ \\

v$_r$ (km s$^{-1}$) 
& $12.53^{+0.01}_{-0.01}$ 
& $5.87^{+0.05}_{-0.05}$ \\

T$_\text{kin}$ (K) 
& $133^{+19}_{-18}$  
& $142^{+103}_{-55}$ \\

T$_\text{rot}$ (K) 
& $74^{+7}_{-7}$ 
& $45^{+3}_{-3}$ \\

\hline

log(N$_{\text{H}_2}$ (cm$^{-2}$)) 
& <22.84 
& <22.55 \\

CO/H$_2$ Ratio 
& >$1.35 \times 10^{-3}$ 
& >$3.09 \times 10^{-5}$ \\

\hline
\end{tabular}
\tablefoot{Best-fit values correspond to the highest log posterior probability with $1\sigma$ uncertainties ($16^\mathrm{th}$/$84^\mathrm{th}$ percentiles of the posterior distributions). The H$_2$ column density (upper limits at the $99.7^\text{th}$ percentile) and CO/H$_2$ ratios assume the H$_2$ shares the same kinetic temperature as the CO.}
\end{table}

\section{Discussion \label{section:4Discussion}}

\subsection{CO/H$_2$ ratios across planet formation: CO depletion to enhancement from protoplanetary to debris disks}
In the previous sections we presented CRIRES+ observations of the stars HD 110058 and HD 131488, which both host edge-on CO-rich exocometary belts. We detected $^{12}$CO rovibrational absorption from the v=2-0 band, and attempted to detect H$_2$ in these systems through its rovibrational v=1-0 S(0) line, arising from absorption of cold molecules in the ground state. Despite the high-SNR achieved on the stars with CRIRES+, and the clear CO detections, we do not detect H$_2$ in either system. To set stringent constraints on its presence, we modelled CO and H$_2$ absorption using the line-by-line spectral modelling tool RADIS to obtain CO column densities and an $\text{H}_2$ column density upper limit. Combined, these yielded $\frac{\text{CO}}{\text{H}_2}$ ratio lower limits of $3.09 \times 10^{-5}$ and $1.35 \times 10^{-3}$ for HD 131488 and HD 110058 respectively. 

\begin{figure}
    \centering
        \includegraphics[width=\linewidth]{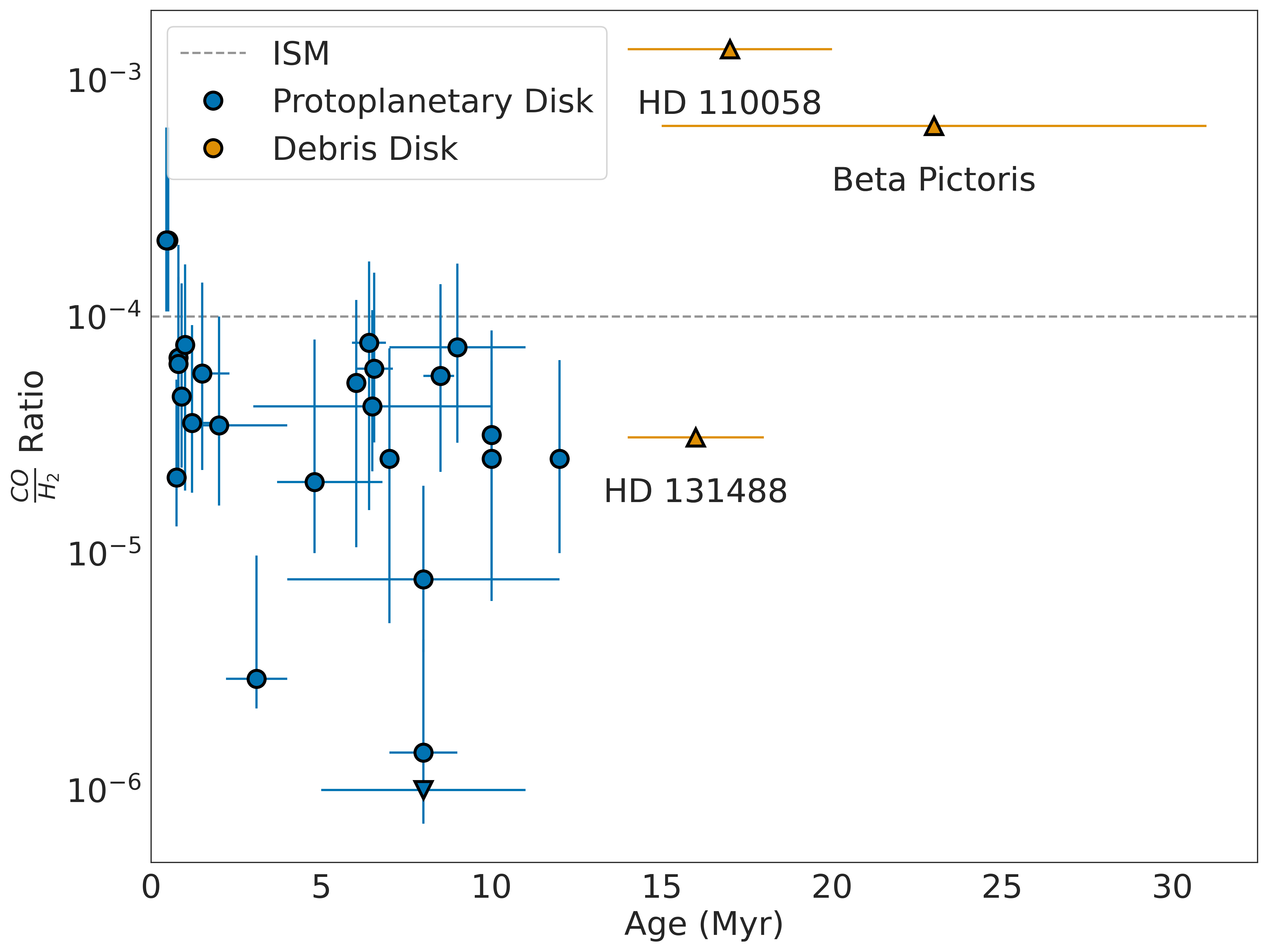}
    \caption{Ratio of CO and H$_2$ abundances for protoplanetary disks, $\beta$ Pictoris, and the two exocometary belts in this study. Details of each point can be found in the references of Table \ref{tab:co_h2_ratios}. This plot has been adapted from and expanded upon using the similar plots in \citet{bergin_determination_2018, zhang_rapid_2020}.}
    \label{fig:col_den_ratios}
\end{figure}

We compare our lower limit for the $\frac{\text{CO}}{\text{H}_2}$ ratio in the exocometary belts' gas to values derived for protoplanetary disks listed in Table \ref{tab:co_h2_ratios} and plotted in Figure \ref{fig:col_den_ratios}. Protoplanetary ratios typically range from $10^{-4}$ to $10^{-6}$ and tend to drop over time within the first 10 Myr of a star's lifetime \citep{favre_significantly_2013, kama_volatile-carbon_2016, mcclure_mass_2016, schwarz_radial_2016, trapman_far-infrared_2017, zhang_mass_2017, bergin_determination_2018, zhang_rapid_2020}. This is in contrast with our first direct measurements (lower limits) for the exocometary belt of HD 110058 which has markedly elevated $\frac{\text{CO}}{\text{H}_2}$ indicating a much more H$_2$-poor environment compared with primordial gas in protoplanetary disks \citep{bergin_determination_2018, zhang_rapid_2020} . The conservative lower limit we obtained via modelling the H$_2$ non-detections in the CRIRES+ K-band spectra of HD 110058 is a factor of 13 larger than the canonical $\frac{\text{CO}}{\text{H}_2}$ ratio of $10^{-4}$ typical of the ISM. For our other target HD 131488, the lower limit still formally allows a primordial  $\frac{\text{CO}}{\text{H}_2}$ ratio similar to that of the ISM and protoplanetary disks. The degree to which an ISM-like model is consistent with the data can also be seen in Figure \ref{fig:hd131488_radis_wide} where the blue model shows the expected line depth for an ISM-like H$_2$ abundance relative the CO column density. It's clear that the spectrum for HD 110058 is not consistent with an ISM-like model whereas the same model for HD 131488 would be consistent with the noise level in our data.

\subsection{H$_2$ + self-shielding of CO: likely insufficient in a primordial scenario}
As discussed in Section \ref{section:1introduction}, another potential way to discriminate between primordial and secondary origin scenarios has been to compare the CO photodissociation timescale to the age of the system, and in particular to evaluate the potential for CO to be shielded from UV radiation. We focus on HD 131488 as we have already shown in Section \ref{subsection:ratios} that HD 110058 has a composition inconsistent with primordial gas. Similar to \citet{marino_exocometary_2016, marino_scattering_2018, matra_detection_2017, matra_ubiquity_2019, brennan_low_2024} we used half our observationally determined line-of-sight column densities of CO and (3$\sigma$ upper limit for) H$_2$ to calculate radial shielding factors \citep{visser_photodissociation_2009} for a CO molecule in the middle of the disk irradiated by the ISRF.

ALMA images of the same CO gas (MacManamon et al. in prep.) show the disk has a vertical height above the midplane of $\sim$ 20 au (where the FWHM of the resolution element for this image was 11 au) and a radial extent of $\sim$ 180 au. The shielding factors of \citet{visser_photodissociation_2009} assume an isotropic radiation field and a spherical cloud shielding a CO molecule in the centre of that cloud. Using the disk's radial column density as the radius of a spherical cloud, we obtain an upper limit of the photodissociation timescale of 55.4 Myr. If, on the other hand, we take the ratio of the height to radius of the disk we can scale the radial column density to estimate a vertical column density. Calculating the shielding provided by a spherical cloud with radius equal to the vertical half height we obtain a lower limit of the photodissociation timescale of 0.11 Myr. The true value is therefore likely between 0.11 and 55.4 Myr; given the disk age of 16 Myr, we cannot formally rule out a primordial origin scenario for HD 131488.

However, given the vertically thin nature of the disk the true value is likely closer to 0.11 Myr and less than the system age, since the column density and therefore the shielding factor averaged over the 4$\pi$ solid angle will be closer to the vertical than the radial value. Additionally, shielding factors are a steep function of CO and H$_2$ column density; a column density that is a factor of just a few lower than our measured radial value shortens the timescale to much less than the system age, rendering shielding much less effective. It is also worth noting that we only take into account the interstellar radiation field, thus neglecting UV radiation from the host star; if significant, this would further shorten the photodissociation timescale. 

Another assumption affecting the calculation is that the CO and H$_2$ column densities measured by CRIRES+ are representative of the disk midplane and thus the bulk of the gas, although both disks' inclinations are not perfectly edge-on (Table \ref{table:stellar_params}). However, the CO in both disks peaks relatively close to the star in both systems, meaning that the bulk of the gas is in a narrow vertical distribution. HD 131488 has a CO inner edge $\lesssim5$ AU and is vertically resolved in recent ALMA data (Mac Manamon et al. in prep.).
Finally, our calculation assumes the photodissociation lifetime is representative of the survival times of CO molecules in the disk in the absence of other destruction/production mechanisms, such as gas-phase chemistry reforming CO from other species.  However, this is only expected to be important in H$_2$-rich environments \citep{higuchi_detection_2017, iwasaki_constraint_2023}, which is unlikely to be the case here given our stringent H$_2$ upper limits. 

Overall, most of these assumptions imply that the CO shielding and in turn photodissociation timescales in HD 131488 calculated from our line of sight column densities of CO and H$_2$, is likely significantly overestimated. Although our quantitative estimate is formally not conclusive, H$_2$ is likely not able to provide sufficient shielding and allow CO to survive over the lifetime (age) of the systems. This would provide further support to the claim that the gas is not primordial but of second generation in HD 131488 as well as in HD 110058, adding to the new conclusive evidence presented here that the CO/H$_2$ ratio differs significantly from protoplanetary disk gas in at least one of our two CO-rich systems. Nevertheless, accurately modelling HD 131488's geometry, local radiation field, shielding from various molecules, outgassing rates, etc. would need to be conducted to determine a more realistic photodissociation timescale.

Although CI shielding has not been considered previously in the context of a primordial scenario it could extend the CO lifespan, though it is unclear how important this is in younger protoplanetary disks where complex chemistry beyond simply photodissociation involving both CO and C is expected. What we find is that in the radial direction the CI column density in \citet{brennan_low_2024} is sufficient to provide shielding over the system lifetime, but it is unclear whether this is possible along the vertical direction as well, where the column density is unconstrained.

For both disks, it is in principle possible that the belts host a hybrid mix of primordial and secondary gas, with their protoplanetary gas having started to dissipate very recently, as suggested by \citet{lisse_spectral_2017}. However, our evidence suggests that even in such a scenario, at least HD 110058 is likely still dominated by second generation gas. Our non-detection of H$_2$ in both systems is consistent with abundance ratios found in Solar System comets \citep{combi_time-dependent_1996}; supporting the hypothesis that the gas in these CO-rich systems is, like the dust, secondary/exocometary in origin rather than primordial.

\subsection{Challenges remain for the second-generation model}
While our results strongly point to a secondary origin for the gas at least for HD 110058, challenges remain for modelling secondary gas production and survivability in exocometary belts. Producing the observed column densities in CO-rich belts through exocometary release without shielding may require unreasonably large gas release rates to counter the destruction timescale of unshielded CO, which is $\sim$ 130 years in the ISRF \citep{heays_photodissociation_2017}. To produce the CO mass of $4-8 \times 10^{-2} M_{\bigoplus}$ found in HD 21997 for example, \citet{kospal_alma_2013} estimates that approximately 6000 Hale-Bopp sized comets would need to be destroyed yearly. CI (as produced through CO photodissociation) has been posited as a shielding agent to allow for a slower CO release rate and longer accumulation timescales \citep{matra_exocometary_2017, marino_population_2020}.

However, \citet{brennan_low_2024} detected both CO and CI in absorption in the same two systems as this study, finding that the observed column densities of CI are much lower than predicted by the exocometary release model of \citet{marino_population_2020} and do not provide sufficient shielding for CO. A similar conclusion was reached by \citet{cataldi_primordial_2023}, who also determined for 14 exocometary belts that the model of \citet{marino_population_2020} over predicts the amount of CI. These two results suggest that either some of the assumptions are incorrect, the second generation model lacks some important physics/chemistry, or that the gas is primordial; our work deems the latter option to be unlikely. In conclusion, while our results provide the first direct evidence for likely insufficient amounts of H$_2$ in CO-rich planetesimal belts to shield CO and a markedly different composition to primordial gas in at least one system, current models of second-generation gas release are not yet able to provide a fully self-consistent explanation, and more modelling work and compositional constraints are needed.

Indeed, the molecular composition of gas in CO-rich exocometary belts also remains an open question. Constraints for the presence of molecules have been placed on $\beta$ Pictoris \citep{matra_molecular_2018}, 49 Ceti \citep{klusmeyer_deep_2021}, HD 21997, HD 121617, HD 131488, and HD 131835 \citep{smirnov-pinchukov_lack_2022}. The upper limits found for molecular species in these works differ from the younger protoplanetary disks for which similar searches have been conducted \citep{smirnov-pinchukov_lack_2022}. The limits are consistent with the secondary generation scenario \citep{matra_molecular_2018} but in the case of CN indicate that CO could be preferentially shielded compared to other molecules \citep{klusmeyer_deep_2021}. Higher signal-to-noise ratio observations are likely needed to find molecules other than CO in gas rich exocometary belts. This unprecedented deep search for molecular hydrogen demonstrates the usefulness of IR instruments like CRIRES+ for constraining the composition of exocometary gas in edge-on exocometary belts. In particular, if the gas is exocometary as strongly supported by our H$_2$ constraints, observations of IR lines in absorption and mm lines in emission can be used to probe the molecular, volatile content of exocometary material. This is of interest for systems such as HD 131488 and HD 110058 which are a few 10s of Myrs-old and are therefore in the latest stages of terrestrial planet formation when volatile delivery to these planets is most likely to occur \citep[and references therein]{morbidelli_building_2012}.

\section{Conclusions} \label{section:5conclusions}
In this work, we presented CRIRES+ spectra of the edge-on exocometary belts around the $\sim$15 Myr-old A stars HD 131488 and HD 110058. We searched for the H$_2$ v=1-0 S(0) rovibrational absorption line at 2223 nm and the $^{12}$CO v=2-0 rovibrational band originating from the ground state of the molecule and therefore probed for cold H$_2$/CO gas within the exocometary belts, along the line of sight to the central stars. We report the following findings: 
\begin{itemize}
    \item We report detections of the $^{12}$CO v=2-0 rovibrational transitions in HD 131488 and HD 110058 and calculate best-fit, line of sight log column densities of 18.05 cm$^{-2}$ and 19.97 cm$^{-2}$ respectively. 
    \item We do not detect absorption from the ground state of the H$_2$ molecule. Modelling this non-detection we calculated upper limits on the $\text{H}_2$ line-of-sight column density of < $10^{22.55}\ \text{cm}^{-2}$  for HD 131488 and  < $10^{22.84}\ \text{cm}^{-2}$ for HD 110058. 
    \item Combined with our CO detections, we compare the $\frac{\text{CO}}{\text{H}_2}$ ratio lower limits (assuming the CO and H$_2$ are at the same temperature) of $3.09 \times 10^{-5}$ (HD 131488) and $1.35 \times 10^{-3}$ (HD 110058) with protoplanetary disks and conclude that the composition of HD 110058 is significantly different to primordial gas found in protoplanetary disks while that of HD 131488 remains potentially consistent with younger disks.
    \item We calculate conservative upper and lower limits for the photodissociation timescale of the CO in HD 131488, finding that we cannot formally rule out a primordial origin scenario for the gas with a simple shielding model. Given the conservative assumptions, we postulate that a more realistic shielding model is likely to rule out a primordial origin scenario. Further observations would be needed to confirm whether HD 131488 has a primordial or secondary composition.
\end{itemize}

\begin{acknowledgements}
  
    KDS and LM acknowledge and thank the Irish research Council (IRC) for funding this work under grant number IRCLA-2022-3788 and the European Union through the E-BEANS project (grant number 100117693). This research used observations made with the European Southern Observatory's CRIRES+ instrument on UT3 of the Very Large Telescope as part of Program ID 111.255A.001. We would like to thank Carlo Manara and the ESO User Support team for their help reducing this data. A. M. H. acknowledges support from the National Science Foundation under Grant No. AST-2307920. AB acknowledges research support by the Irish Research Council under grant GOIPG/2022/1895.
    
\end{acknowledgements}

\section*{Data availability}
The observations detailed in this publication are publicly available in the ESO Science Archive Facility (http://archive.eso.org) under the program ID 111.255A.001. Data products will be shared on reasonable request to the corresponding author.

\bibliographystyle{aa.bst} 
\bibliography{hd131488_crires}

\begin{appendix}

\section{Molecfit telluric corrections}
\label{appendix:molecfit}

\begin{figure} [H]
    \centering
    \includegraphics[width=\linewidth]{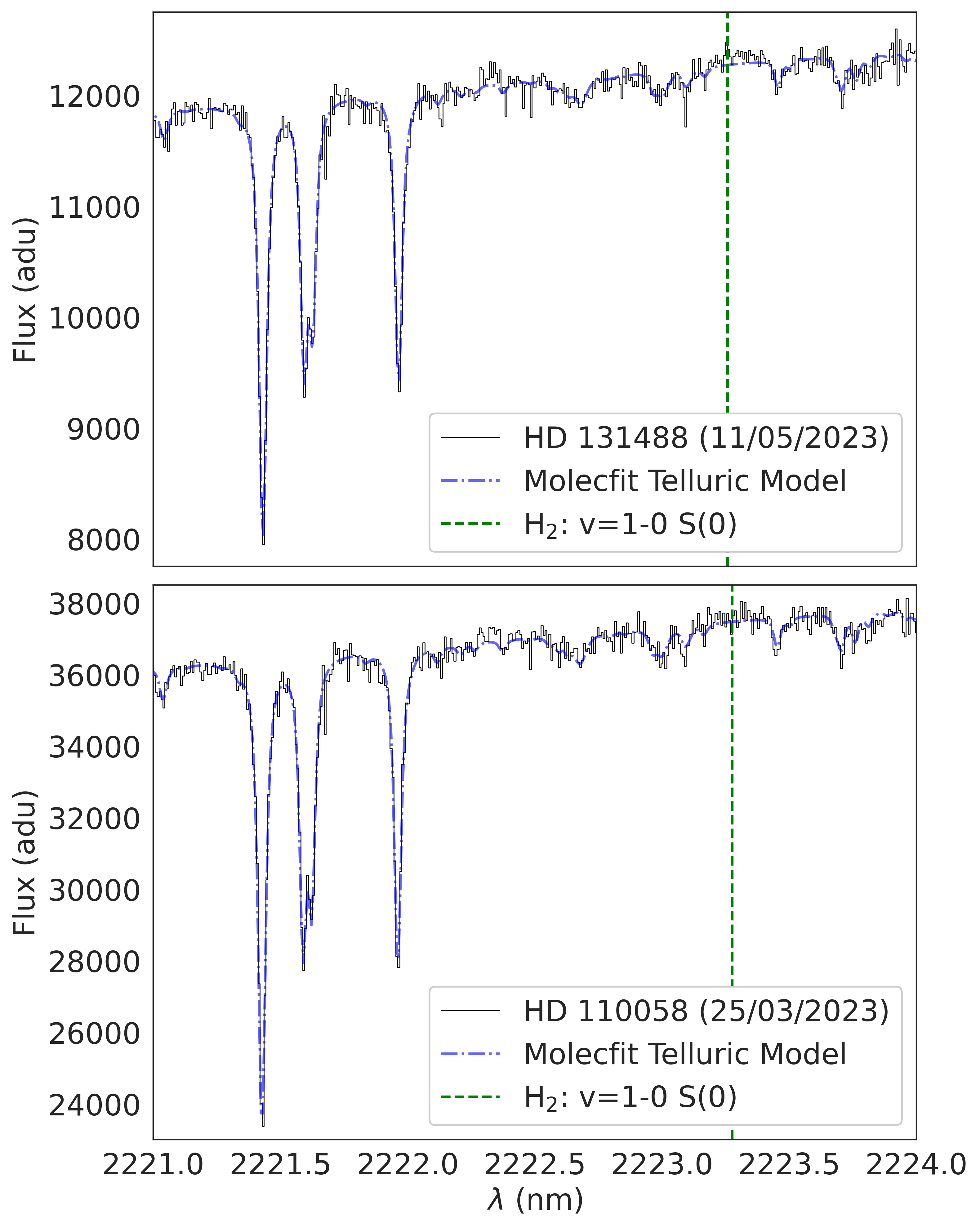}
    \caption{CRIRES+ spectra of HD 131488 and HD 110058 near the H$_2$ v=1-0 S(0) line (black lines). In blue is the best-fit atmospheric transmission model from \texttt{molecfit} consisting of CH$_4$ and H$_2$O absorption lines, combined with our (linear) best-fit continuum model. The vertical green line denotes the expected location of the H$_2$ transition of interest.}
    \label{fig:molecfot}
\end{figure}

\begin{figure} [H]
    \centering
    \includegraphics[width=\linewidth]{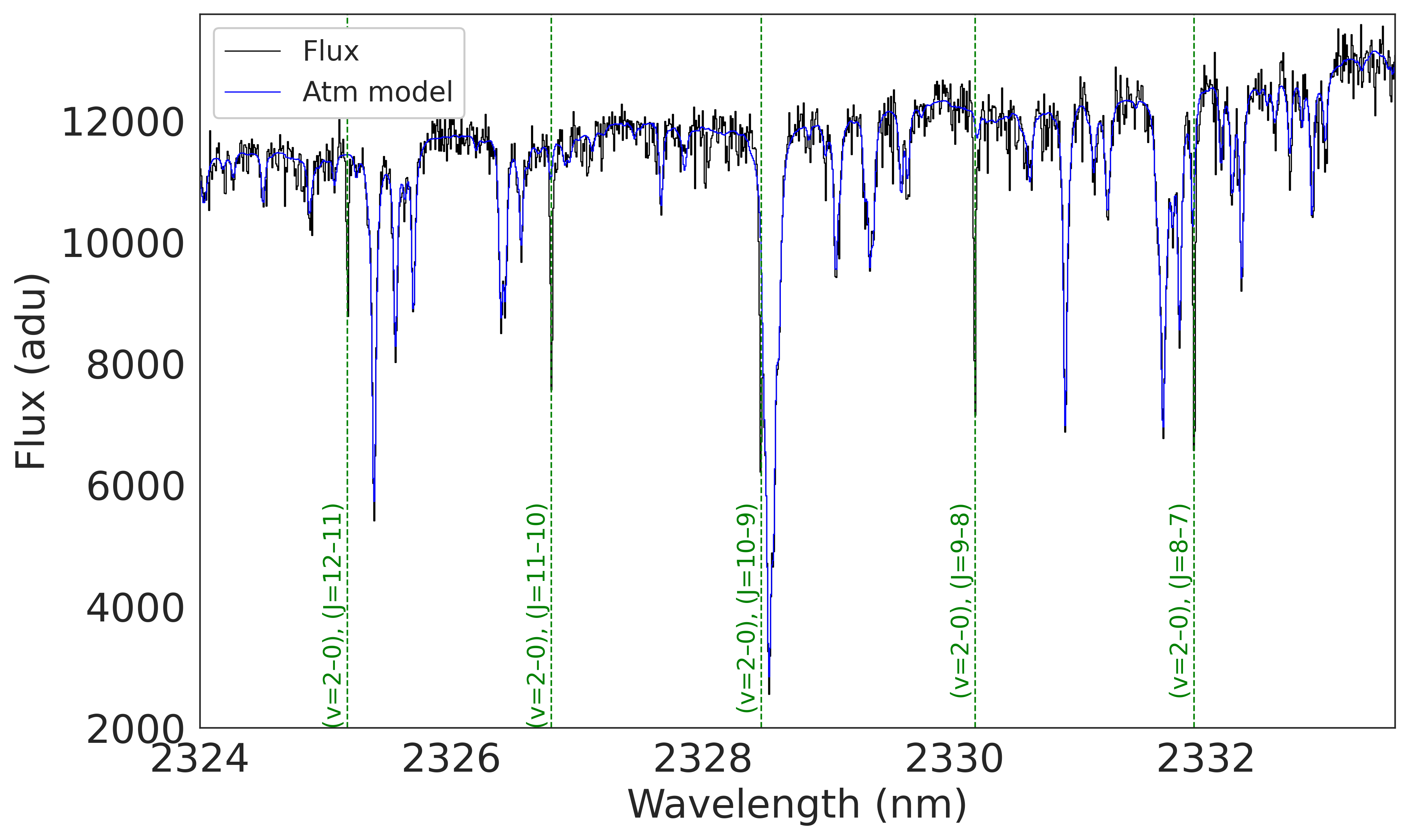}
    \caption{A portion of an example spectrum (black) overlaid with the telluric fit (blue) produced by molecfit over the data from a single nodding position/day combination (the combined nodding A frames on the night of 08/03/2024). The locations of $^{12}$CO lines are marked with green dashed lines.}
    \label{fig:12CO_telluric}
\end{figure}

\section{On CRIRES+ wavelength calibration offset between nodding positions}
\label{appendix:superresolution}
\begin{figure} [H]
    \centering
    \includegraphics[width=1\linewidth]{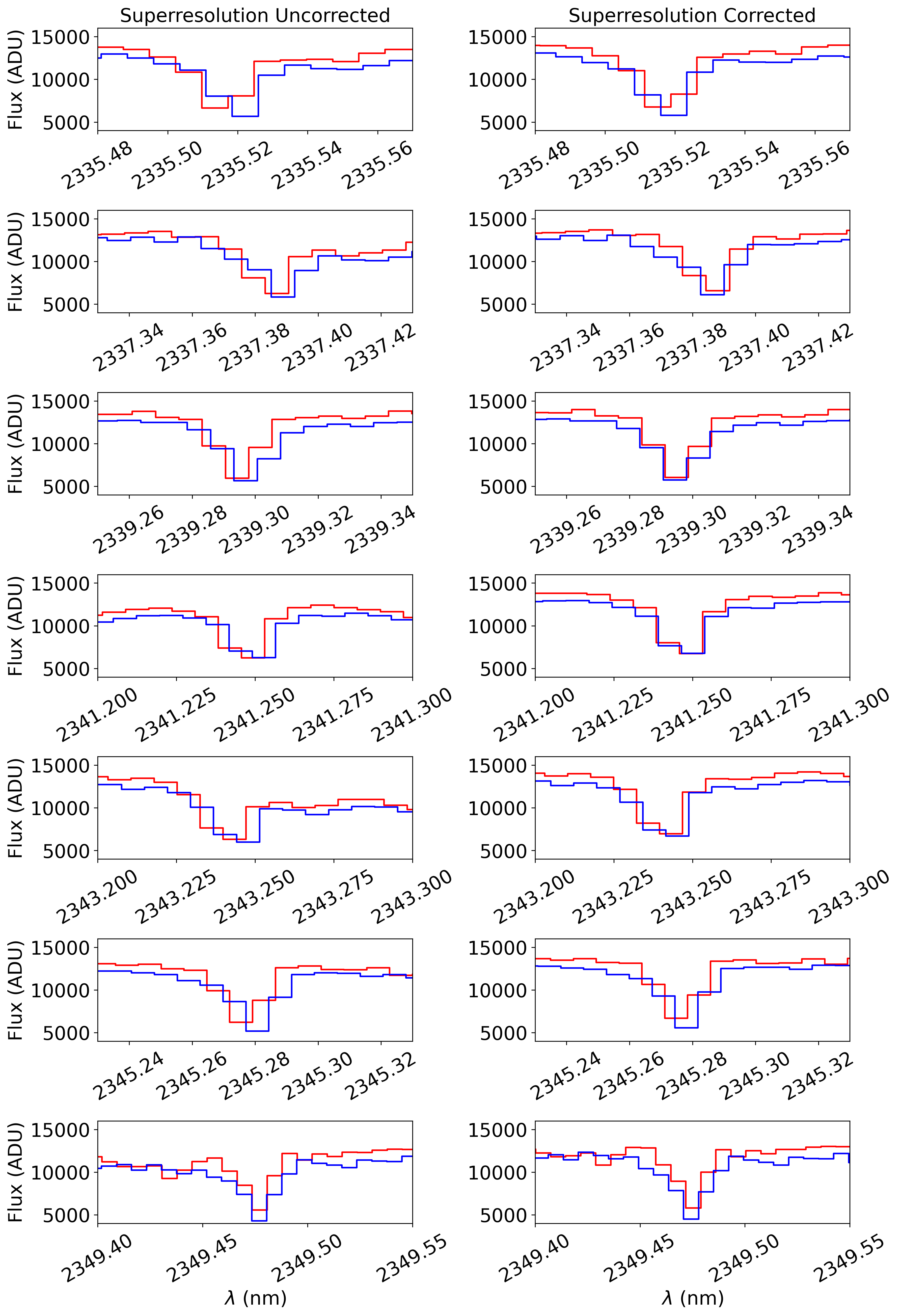}
    \caption{$^{12}$CO lines before and after superresolution correction. In red, the nodding A frames should align with the nodding B frames in blue as these observations were taken on the same day if the wavelength solution is accurate.}
    \label{fig:superres_demo}
\end{figure}

\begin{figure} [H] 
    \centering
    \includegraphics[width=1\linewidth]{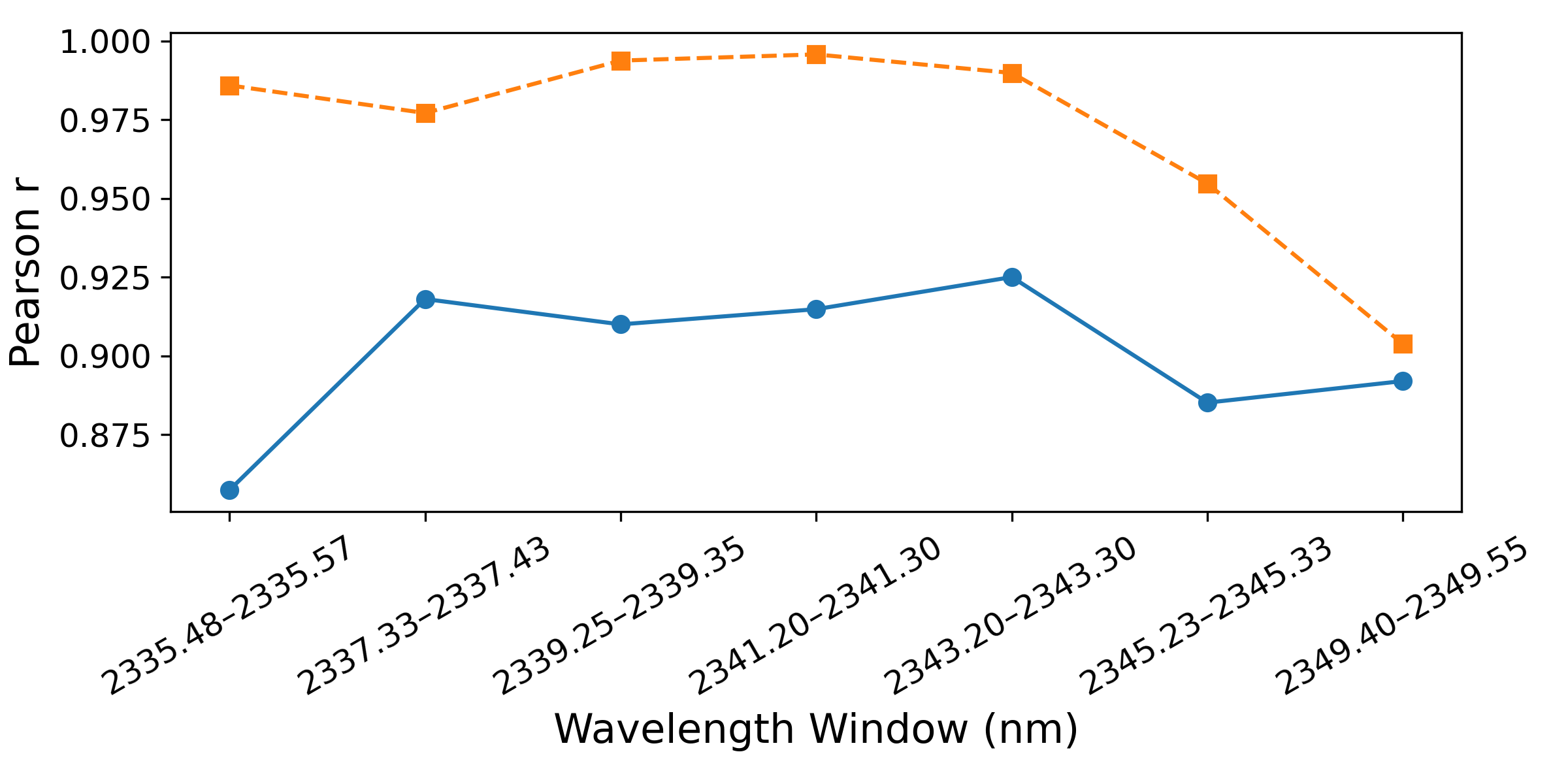}
    \caption{Pearson correlations of the nodding A and B positions for the uncorrected (blue) and post-correction (orange) $^{12}$CO lines from Figure \ref{fig:superres_demo}.}
    \label{fig:superres_corr}
\end{figure}

Our observations are taken with the 0.2" width slit on CRIRES. When reducing our data, the ESO data reduction pipeline reported that for both of our targets, the width of the slit was not fully illuminated. This has two consequences for our observations detailed in the ESO CR2RES pipeline user manual in the known issues section 7.1 on "Superresolution". Relevant to this study, this gives us increased resolution but at the cost of the default wavelength solution produced by the ESO pipeline being unreliable between nodding frames. To account for this, we take the combined spectra for each nodding position and separately feed each one through molecfit to acquire a final wavelength solution fit to the tellurics in the data. For each star this means we fit 8 spectra for $^{12}$CO (2 days, 2 nodding positions, and 2 sub-orders) and 4 spectra for H$_2$ (2 days and 2 nodding positions). When presenting the results in this paper, we show plots of the median of these spectra. Any combination or modelling of spectra with an incorrect wavelength solution would distort the line shapes and positions, diminishing the accuracy of our results. The degree to which this is an issue is shown in Figures \ref{fig:superres_demo} and \ref{fig:superres_corr} which show the pre-corrected wavelength solution can err by as much as a pixel. The higher post-correction Pearson correlations of the A and B nodding frames in Figure \ref{fig:superres_corr} demonstrates numerically that the alignment of the spectra is improved.

\section{MCMC posterior/prior distributions}
\begin{figure} [H]
    \centering
    \includegraphics[width=1\linewidth]{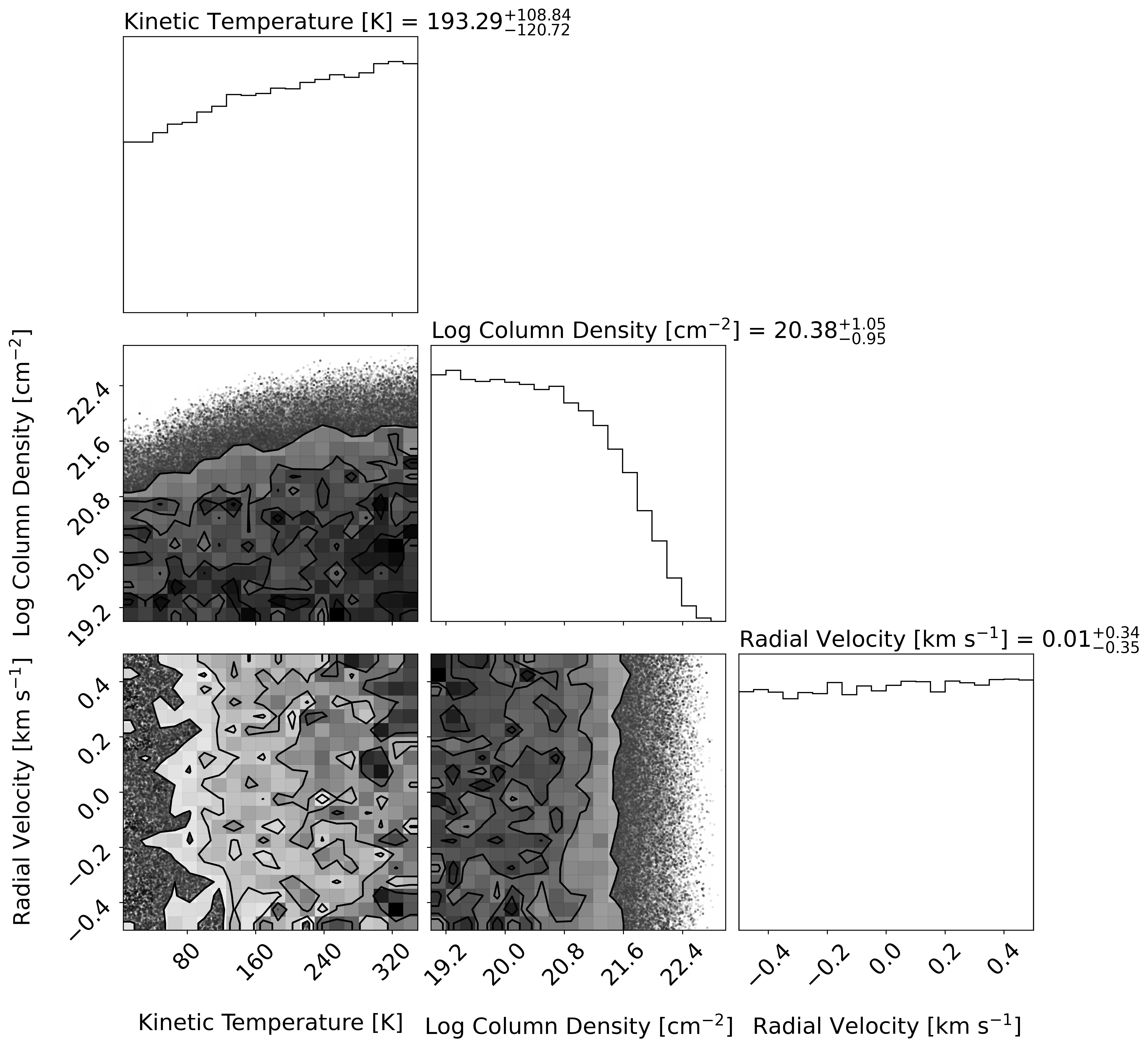}
    \caption{Posterior probability distributions of temperature, column density, and radial velocity from the 2223.29 nm rovibrational H$_2$ absorption model fit to the HD 131488 data. Marginalised distributions for each parameter are displayed on the diagonal. Quantities quoted above each panel are the median and error bars calculated from the 16$^\text{th}$ and 84$^\text{th}$ quantiles.}
    \label{fig:hd131488_posterior}
\end{figure}

\begin{figure} [H]
    \centering
    \includegraphics[width=\linewidth]{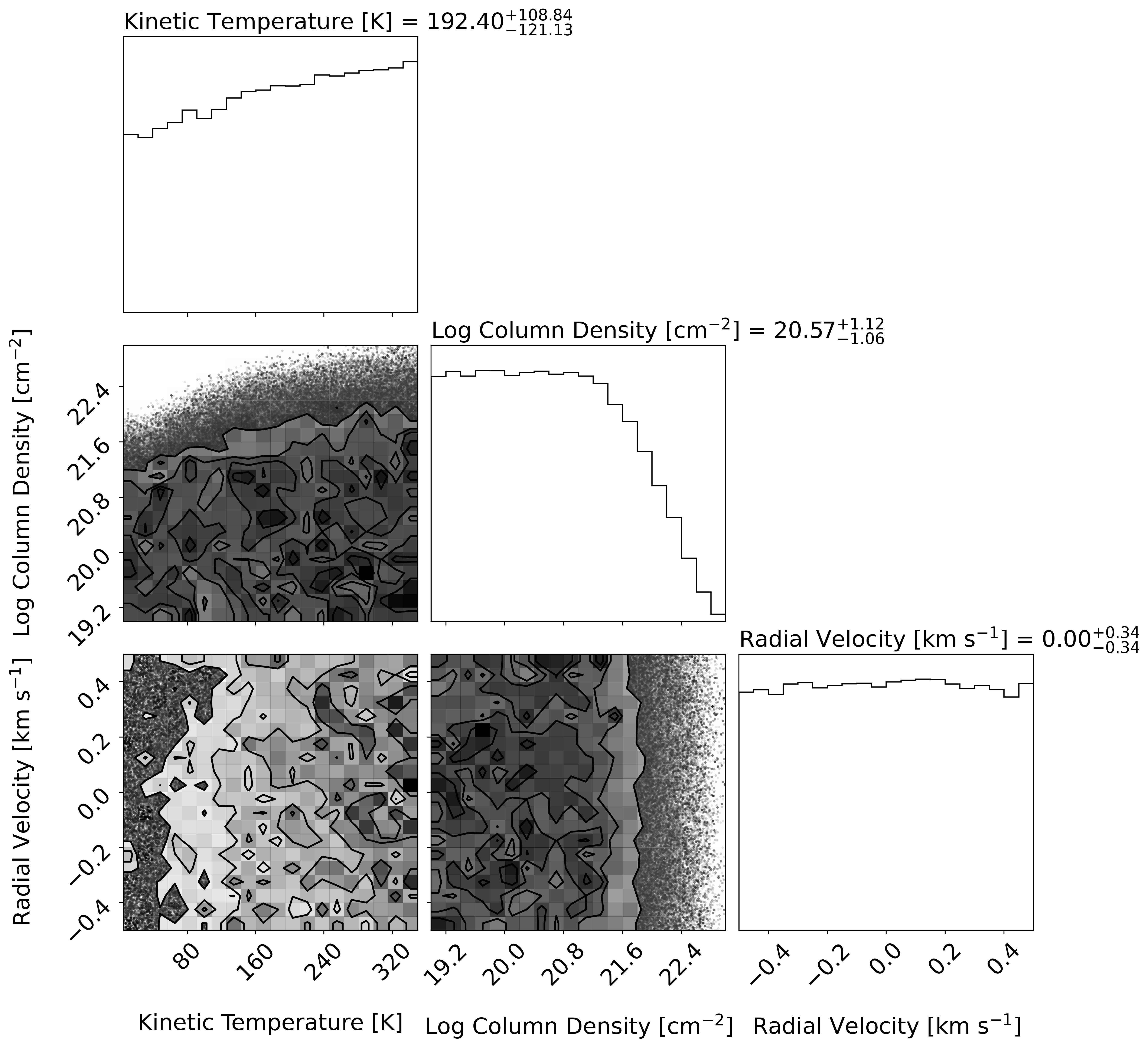}
    \caption{Posterior probability distributions of temperature, column density, and radial velocity from the 2223.29 nm rovibrational H$_2$ absorption model fit to the HD 110058 data. Marginalised distributions for each parameter are displayed on the diagonal. Quantities quoted above each panel are the median and error bars calculated from the 16$^\text{th}$ and 84$^\text{th}$ quantiles.}
    \label{fig:hd110058_posterior}
\end{figure}

Table \ref{tab:MCMC_priors} gives the priors used in our MCMC simulations. In Figures \ref{fig:hd131488_posterior}, and \ref{fig:hd110058_posterior} we show the posterior distributions from the MCMC simulations for H$_2$ and these are also stated in Table \ref{tab:MCMC_results}. Given that H$_2$ is not detected in either system, the 2d corner plots show upper limits for the H$_2$ column density and unconstrained temperature as expected for a non-detection. Higher temperatures allow for larger column densities as the thermal broadening produces shallower, wider lines which are more difficult to detect than the narrower, deeper lines produced by cold gas. 

 For CO, each prior bound was selected to be much larger than the posterior CO distributions found by \citet{brennan_low_2024}. The H$_2$ column density priors allow the H$_2$ to be the same as the CO or up to $10^5$ times larger (an order of magnitude larger than an ISM-like abundance). This allows for anything between a non-detection and line profiles much larger than those we would expect for a detection. The temperature range for H$_2$ was chosen to be smaller to speed up convergence while still being much larger than the CO posterior temperatures. Smaller column density ranges were used in test runs and found not to significantly influence the upper limits calculated later. We ran each MCMC for 50000 steps with 100 walkers with a burn-in of the first 10\% of the steps and we visually inspected the chains to ensure convergence. 

\begin{table}
\caption{Bounds of the uniform priors for each parameter in the MCMCs.}
\label{tab:MCMC_priors}
\begin{tabular}{c c c} 
\hline\hline
Parameter & CO & H$_2$ \\
\hline
log(N (cm$^{-2}$)) &
$10^{15} \rightarrow 10^{23}$ & $10^{19} \rightarrow 10^{23}$ \\ 
v$_r$ (HD 110058) (km s$^{-1}$) &
$10 \rightarrow 15$ & CO best fit $\pm 0.5$ \\
v$_r$ (HD 131488) km s$^{-1}$ &
$2 \rightarrow 8$ & CO best fit $\pm 0.5$ \\
T$_\text{kin}$ (K) & $5\rightarrow700$ & $5\rightarrow350$ \\
T$_\text{rot}$ (K) &
$5\rightarrow700$& NA \\ 
\hline
\end{tabular}
\end{table}

\label{appendix:MCMC}

\begin{figure} [H]
    \centering
    \includegraphics[width=1\linewidth]{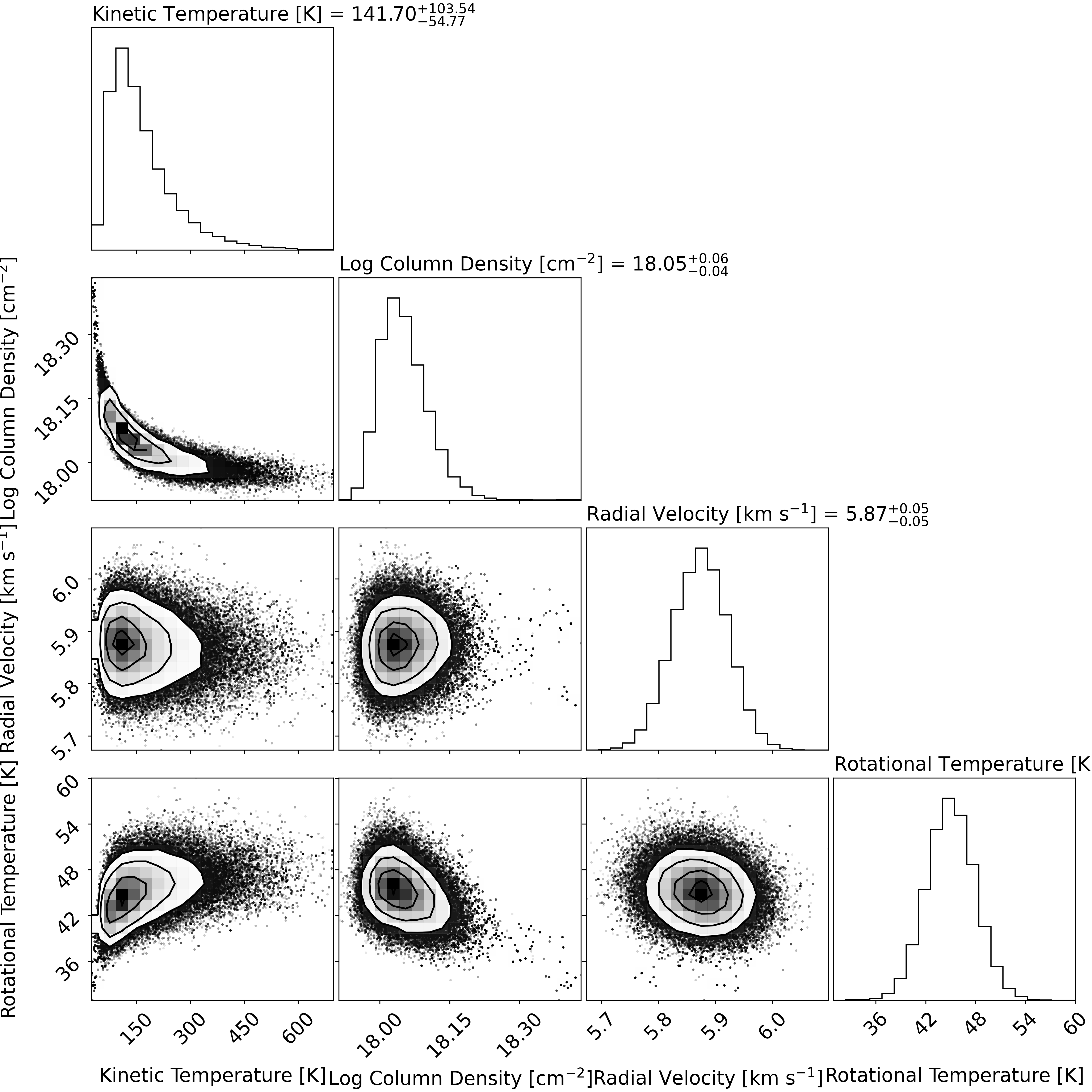}
    \caption{Posterior probability distributions of excitation temperature, kinetic temperature, radial velocity, and column density from the for the $^{12}$CO absorption model fit to the HD 131488 data. Marginalised distributions for each parameter are displayed on the diagonal. Quantities quoted above each panel are the median and error bars calculated from the 16$^\text{th}$ and 84$^\text{th}$ quantiles.}
    \label{fig:hd131488_CO_posterior}
\end{figure}

\begin{figure} [H]
    \centering
    \includegraphics[width=\linewidth]{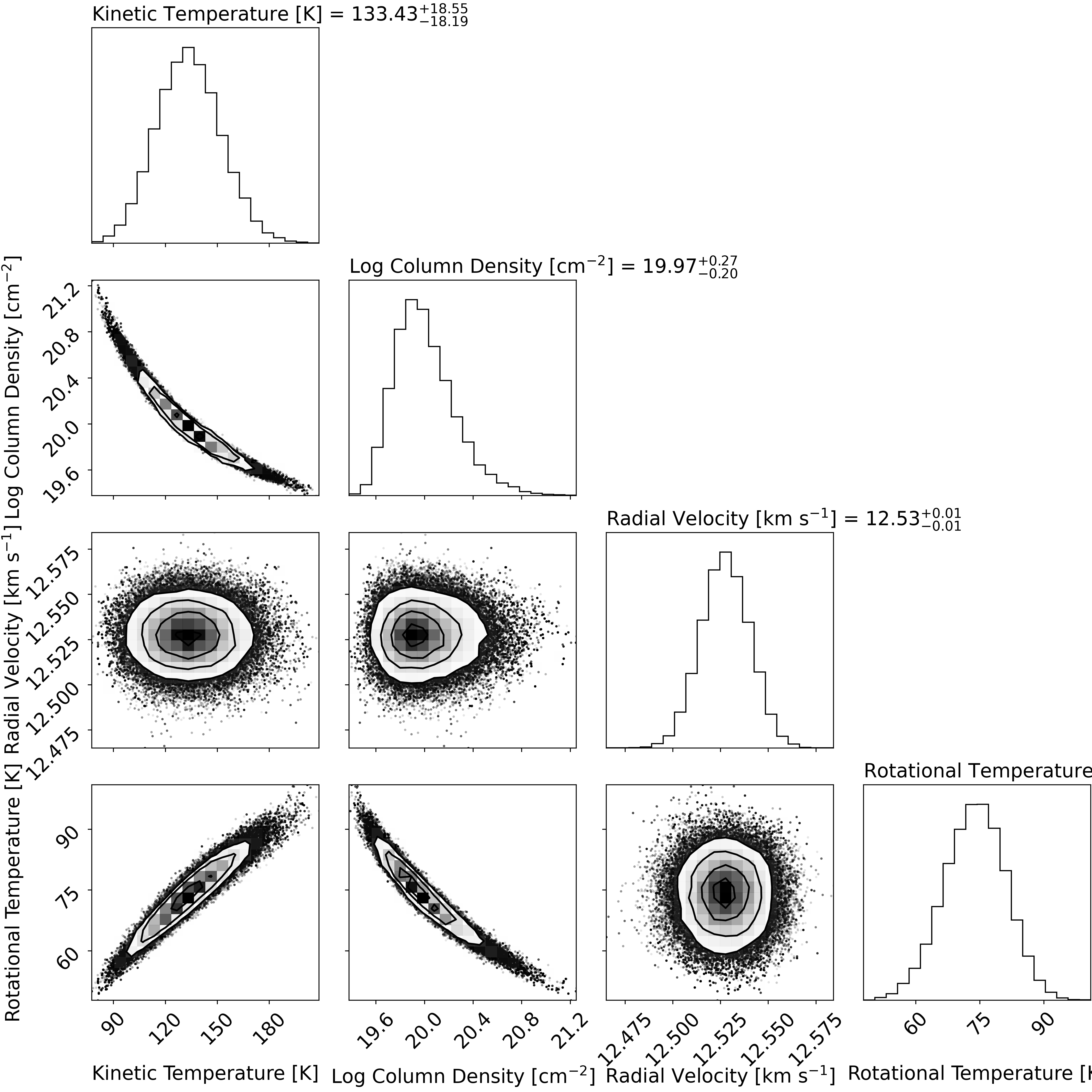}
    \caption{Posterior probability distributions of excitation temperature, kinetic temperature, radial velocity, and column density from the for the $^{12}$CO absorption model fit to the HD 110058 data. Marginalised distributions for each parameter are displayed on the diagonal. Quantities quoted above each panel are the median and error bars calculated from the 16$^\text{th}$ and 84$^\text{th}$ quantiles.}
    \label{fig:hd110058_CO_posterior}
\end{figure}

\section{CO/H$_2$ ratios for protoplanetary and exocometary belt gas} \label{subsection:ratios}
\label{appendix:ppd_literature}
\raggedbottom
The determination of the ages and gas masses of protoplanetary disks is difficult and an active area of research \citep{zhang_rapid_2020}. A variety of techniques have been used to determine the gas mass including assuming a gas-to-dust ratio, and assuming ISM-like CO isotopologue ratios to give estimates of the mass of the typically optically thick $^{12}$CO in disks. The amount of H$_2$ can be estimated from assuming ISM-like HD to H$_2$ ratios, dust-to-gas ratios, gas/dust kinematics, or through N$_2$H$^+$. 

\begin{landscape}
\begin{table}
    \caption{CO to H$_2$ Ratios and Age for Disks}
    \label{tab:co_h2_ratios}
    \centering
    \begin{tabular}{l c c c c c c c}
        \hline\hline
        Name & $\frac{\text{CO}}{\text{H}_2}$ Ratio & Age (Myr) & CO Method & H$_2$ Method & Age Method & Type & References \\
        \hline
        ISM & $-4.00$ & -- & -- & -- & -- & ISM & -- \\
        HL Tau & $-4.17^{+0.04}_{-0.03}$ & $0.8^{+0.25}_{-0.25}$ & $^{13}$CO & Dust & SED & PPD & [1,2]\\
        DG Tau & $-4.34^{+0.04}_{-0.04}$ & $0.9^{+0.25}_{-0.25}$ & $^{13}$CO & Dust & SED & PPD & [1,2] \\
        TMC1A & $-3.68^{+0.02}_{-0.04}$ & $0.5^{+0.1}_{-0.1}$ & $^{13}$CO & Dust & SED & PPD & [1,2] \\
        Oph-IRS67 & $-3.68^{+0.30}_{-0.30}$ & $0.45^{+0.1}_{-0.1}$ & C$^{17}$O & Dust & SED & PPD & [3,2] \\
        DM Tau & $-4.24^{+0.15}_{-0.21}$ & $1.5^{+0.8}_{-0.25}$ & $^{13}$CO, C$^{18}$O & N$_2$H$^+$ & SED & PPD & [2,4,30] \\
        TW Hya & $-5.84^{+0.00}_{-0.30}$ & $8^{+1}_{-1}$ & $^{13}$CO, C$^{18}$O & Dust & TW Hydra & PPD & [4,28] \\
        HD 163296 & $-4.28^{+0.17}_{-0.21}$ & $6.03^{+0.28}_{-0.27}$ & $^{13}$CO, C$^{18}$O & N$_2$H$^+$ & Isochrones & PPD & [4,5,30] \\
        IM Lup & $-4.45^{+0.21}_{-0.20}$ & $1.2^{+0.8}_{-0.25}$ & $^{13}$CO, C$^{18}$O & N$_2$H$^+$ & Isochrones & PPD & [4,29,30] \\
        J1609 J1608 & $<-6.00$ & $8^{+3}_{-3}$ & C$^{18}$O & N$_2$H$^+$ & Upper Sco & PPD & [6,30] \\
        HD 100546 & $-4.70^{+0.30}_{-0.30}$ & $4.8^{+2}_{-1.1}$ & $^{12}$CO, $^{13}$CO, C$^{18}$O & HD & Isochrones & PPD & [7,8,9,31] \\
        PDS 66 & $-5.53^{+0.28}_{-0.29}$ & $3.1^{+0.9}_{-0.9}$ & $^{13}$CO, C$^{18}$O & N$_2$H$^+$ & Isochrones & PPD & [10,30,32] \\
        AA Tau & $-4.60^{+0.24}_{-0.22}$ & $7$ & $^{13}$CO, C$^{18}$O & N$_2$H$^+$ & Isochrones & PPD & [11,30,32] \\
        AS 209 & $-4.68^{+0.25}_{-0.26}$ & $0.75^{+0.25}_{-0.25}$ & $^{13}$CO, C$^{18}$O & N$_2$H$^+$ & $\rho$ Ophiuchi & PPD & [12,30, 33] \\
        CQ Tau & $-4.50^{+0.30}_{-0.08}$ & $10.0$ & $^{13}$CO, C$^{18}$O & N$_2$H$^+$ & Isochrones & PPD & [13,30] \\
        Elias 2-27 & $-4.20^{+0.14}_{-0.19}$ & $0.8$ & $^{13}$CO, C$^{18}$O & N$_2$H$^+$ & $\rho$ Orph & PPD & [14,30,34] \\
        GM Aur & $-4.38^{+0.21}_{-0.23}$ & $6.5^{+3.5}_{-3.5}$ & $^{13}$CO, C$^{18}$O & N$_2$H$^+$ & Isochrones & PPD & [15,30,35] \\
        HD 135344B & $-4.13^{+0.10}_{-0.17}$ & $9.0^{+2.0}_{-2.0}$ & $^{13}$CO, C$^{18}$O & N$_2$H$^+$ & Isochrones & PPD & [16,30] \\
        HD 143006 & $-5.11^{+0.25}_{-0.23}$ & $8.0^{+4.0}_{-4.0}$ & $^{13}$CO, C$^{18}$O & N$_2$H$^+$ & Upper Sco & PPD & [17,30] \\
        HD 34282 & $-4.11^{+0.07}_{-0.12}$ & $6.4^{+0.5}_{-0.5}$ & $^{13}$CO, C$^{18}$O & N$_2$H$^+$ & Isochrones & PPD & [18,30] \\
        LkCa 15 & $-4.46^{+0.20}_{-0.21}$ & $2.0^{+2.0}_{-1.0}$ & $^{13}$CO, C$^{18}$O & N$_2$H$^+$ & Isochrones & PPD & [19,30] \\
        HD 31648 & $-4.22^{+0.15}_{-0.21}$ & $6.55^{+0.55}_{-0.55}$ & $^{13}$CO, C$^{18}$O & N$_2$H$^+$ & Isochrones & PPD & [20,30] \\
        HD 36112 & $-4.25^{+0.15}_{-0.21}$ & $8.5^{+0.4}_{-0.5}$ & $^{13}$CO, C$^{18}$O & N$_2$H$^+$ & Isochrones & PPD & [21,30] \\
        RXJ 1615.3-3255 & $-4.12^{+0.08}_{-0.12}$ & $1.0$ & $^{13}$CO, C$^{18}$O & N$_2$H$^+$ & Lupus & PPD & [22,30] \\
        RXJ 1842.9-3532 & $-4.60^{+0.21}_{-0.22}$ & $10.0$ & $^{13}$CO, C$^{18}$O & N$_2$H$^+$ & Isochrones & PPD & [23,30] \\
        V4046 Sgr & $-4.60^{+0.20}_{-0.21}$ & $12.0$ & $^{13}$CO, C$^{18}$O & N$_2$H$^+$ & Isochrones & PPD & [24,30] \\
        HD 131488 & $>-4.66$ & $16^{+2}_{-2}$ & $^{12}$CO & H$_2$ & Upper Centaurus Lupus & Debris Disk & [25,26] \\
        HD 110058 & $>-2.45$ & $17^{+3}_{-3}$ & $^{12}$CO & H$_2$ & Lower Centaurus Crux & Debris Disk & [25,27] \\
        Beta Pictoris & $>-3.19$ & $23^{+8}_{-8}$ & $^{12}$CO & H$_2$ & $\beta$ Pictoris & Debris Disk & [36,37] \\
        \hline
    \end{tabular}
    \tablebib{
       (1) \citet{zhang_rapid_2020}; (2) \citet{robitaille_interpreting_2007}; 
        (3) \citet{artur_de_la_villarmois_chemistry_2018}; (4) \citet{zhang_systematic_2019}; (5) \citet{wichittanakom_accretion_2020}; 
        (6) \citet{anderson_probing_2019}; (7) \citet{kama_volatile-carbon_2016}; (8) \citet{bruderer_warm_2012}; (9) \citet{pineda_high-resolution_2019};         (10) \citet{asensio-torres_perturbers_2021}; (11) \citet{schneider_multi-epoch_2018}; (12) \citet{natta_accretion_2006}; (13) \citet{mannings_high-resolution_2000}; (14) \citet{andrews_protoplanetary_2009}; (15) \citet{macias_multiple_2018}; (16) \citet{muller_hd_2011}; (17) \citet{pecaut_revised_2012}; (18) \citet{merin_study_2004}; (19) \citet{kraus_coevality_2009}; (20) \citet{simon_dynamical_2000}; (21) \citet{vioque_gaia_2018}; (22) \citet{makarov_lupus_2007}; (23) \citet{neuhauser_search_2000}; (24) \citet{torres_young_2008}; (25) This work; (26) \citet{pecaut_revised_2012}; (27) \citet{pecaut_revised_2012};(28) \citet{rhee_characterization_2007}; (29) \citet{mawet_direct_2012}; (30) \citet{trapman_exoalma_2025}; (31) \cite{bergin_determination_2018}; (32) \citet{siess_internet_2000}; (33) \citet{dantona_evolution_1997}; (34) \citet{luhman_low-mass_1999}, (35) \citet{zaire_magnetic_2024} (36) \citet{lee_revisiting_2024}; (37) \citet{lecavelier_des_etangs_deficiency_2001}
    }
\end{table}
\end{landscape}

\end{appendix}

\end{document}